\documentclass[aps,prd,onecolumn,groupedaddress,showpacs,nofootinbib,amssymb]{revtex4} 
\usepackage[T1]{fontenc}
\usepackage[latin1]{inputenc}
\usepackage{graphicx}
\usepackage[english]{babel}
\usepackage{graphicx}
\usepackage{bm}
\usepackage{amsmath}
\usepackage{amssymb}
\usepackage{amsfonts}
\usepackage{epsfig}
\usepackage{colordvi}
\usepackage{color}

\begin{document}
\title{Probing Long Gamma Ray Bursts  progenitor mass by Gravitational Waves}

 \author{Mariafelicia De Laurentis$^{1,3}$\footnote{e-mail address: mfdelaurentis@tspu.edu.ru}, Fabio Garufi$^{2,3}$\footnote{garufi@na.infn.it},
          Maria Giovanna Dainotti$^{4,5,6}$\footnote{maria.dainotti@riken.jp},
        Leopoldo Milano$^{2,3}$\footnote{milano@na.infn.it}
          }

   \affiliation{$^{1}$Tomsk State Pedagogical University, ul. Kievskaya, 60, 634061 Tomsk, Russian Federation}
   
    \affiliation{$^{2,3}$Dipartimento di Fisica, Universit\`a di Napoli "Federico II" Compl. Univ. di
Monte S. Angelo, Edificio G, Via Cinthia, I-80126, Napoli, Italy\\
   INFN sez. di Napoli Compl. Univ. di Monte S. Angelo, Edificio G, Via Cinthia, I-80126 - Napoli, Italy}
             
            \affiliation{$^{4,5,6}$Astrophysical Big Bang Laboratory, RIKEN, 2-1, Hirosawa, Wako-shi, Saitama 351-0198, Japan \\
             Department of Physics \& Astronomy, Stanford University, Via Pueblo Mall 382, Stanford, CA 94305-4060 \\ 
             Astronomical Observatory, Jagiellonian University, Krakow, Poland} 
						
						 %\date{\today}

\begin{abstract}% %  
In this work we present a procedure to infer the mass of progenitors and remnants of Gamma Ray Bursts (GRB), starting from the observed  energy $E_{iso}^{GRB}$ emitted isotropically and considering the associated emission of Gravitational Waves (GW) $ E_{iso}^{GW}$ in the different phases.
 We assume that the GW energy of the progenitor $E_{PROG}^{GW}$ is emitted partially during a star collapse, and the residual energy is related to the GW energy emitted by the remnant.
 
 We take a sample of $237$ Long GRB, and use an hybrid Montecarlo procedure to explore, for each of them, a  region of possible solutions of $ E_{iso}^{GW}$ as a function of the masses, radii, oblateness, rotation frequencies of progenitor and remnant and the fraction of energy $k$ emitted as GW by the GRB.
 
 We discriminate between a Neutron Star (NS) or Black Hole (BH) for the remnant and obtain interesting values for the GW emitted by the remnant NS or BH, for the conversion factor $k$ of  and for the masses and radii of GRB progenitor stars. We also obtain remnant populations with mean masses, mean GW frequencies and GRB frequency of GW emission in agreement with the most accepted models.
\end{abstract}


\maketitle

%%%%%%%%%%%%%%%%%%%%%%%%%%%
\section{Introduction}
%%%%%%%%%%%%%

 The leading theories for the origin of long GRB (LGRB)  assume that the progenitors are massive stars, such as for example Wolf-Rayets (WR) and red giant, collapsing to a compact object (black hole (BH) or neutron star (NS)) \cite{Woosley}. 
Wolf-Rayet stars are candidate progenitors for type Ib/Ic supernovae (SNe), which give birth to either a NS or a BH \cite{Smartt 2009}. They may be the progenitors of long and soft GRBs, or at least part of them \cite{Woosley2011}. The question is whether WRs collapse into a BH, or give birth to a type Ib or type Ic SN. This is relevant, because in this paper we aim to classify the GRB observations according to their progenitor mass. In the scenario of possible GRB progenitors, type Ib, Ic and Ib/c supernovae (SNe) or hypernovae are produced \cite{Davies,Georgy2009}. It is realistic that core-collapse SNe induce LGRB signals when stellar cores undergo fast rotations before the collapse. The bulk of matter then forms a torus around the newly-formed central compact object and, following the accretion of the torus, feeds the GRB \cite{Paczynski,Fryer}. As it is well known, the first non-electromagnetic signal that can be related by a GRB is a gravitational radiation emission in the form of GWs. Such perturbations are not generated by the burst itself, but by the progenitor through the central engine \cite{Kobayashi}.
Furthermore, in the case of LGRBs, the GW signal should arrive soon before the burst, {\it i.e.} during the core-collapse phase or during the production of the central BH which could be originated also by  merging processes. For the SNe-GRB connection, several papers have been written; a comprehensive discussion on the problem can be found in \cite{Woosley1} and references therein. Considering progenitors, particularly in the case of LGRBs, there are no final models for the GWs emission but several remarkable papers have been published pointing out the problem \cite{LIGO,Fryer1,Kobayashi,Meszaros,Meszaros1,vanPutten,Ott}. The expected signal would lie in the range $10^2\div10^3$ Hz and, in principle, could be detected only for events up to a few tens of Mpc \cite{Meszaros}. The limitation in distance is the main and critical problem for the detection of possible GW signals related to GRB progenitors. In this paper we suppose that the progenitor masses, at the origin of LGRBs, could be related to the compact objects resulting as remnants of SNe or collapsar processes. The resulting remnant can be either a NS or BH objects taking into account the emitted energies as GW in the GRB prompt and in the afterglow phase. 
The GW production can give hints on the total mass of the collapsing system by estimating the global emitted gravitational energy. 
By relating the energy emitted in GW by the progenitor to a fraction $k$ of the electromagnetic GRB prompt emission energy, and the energy emitted in GW by the remnant, one could trace the progenitor mass. 

The main steps that we will follow are:
\begin{itemize}
\item We assume that the energy released by the GRB progenitor during the collapse is partially converted into GWs, {\it i.e.} we assume that the order of magnitude of gravitational energy is a certain fraction $k$ of the energy emitted in the electromagnetic $E_{iso}^{GRB}$; 
\item  As a second step, we constrain the progenitor masses assuming that the energy balance of progenitor in the collapsar phase will be: $E_{PROG}^{GW} = E_{iso}^{GW}+ E_{RMN}^{GW}$, where $E_{PROG}^{GW}$ is the energy emitted by the progenitor as GW's, $E_{iso}^{GW}$ is the energy converted in GW during the GRB prompt, and finally $E_{RMN}^{GW}$ is the GW energy emitted by the remnant.
%% FG Edit%%%
We are aware that the energy emitted in GW by the remnant is not in principle directly related to the energy of the progenitor, but this is a rough estimate to correct the total GW emission of the GRB. On the other hand it can be roughly justified by the consideration that some of the energy of the progenitor goes into the kinetic energy of the remnant, {\it e.g.} in rotational energy, and thus in the latter GW emission.
%%% End FG edit %%%
\item We compare the dimensionless amplitude $h_0$ of GWs, during the collapsar phase, with the current sensitivity curves of the ground interferometric GW antennas, like LIGO and VIRGO, and the future interferometric GW antennas, like Advanced LIGO, Advanced VIRGO and the forthcoming Einstein underground interferometric GW antenna (ET) (see \cite{LIGO,VIRGO,ET}); 

   In this context we choose progenitor masses in the interval $10 M_{\odot}\leq M \leq 60 M_{\odot}$, considering the more massive progenitors that can also be $60 M_{\odot}$. Therefore, in this scenario, we take also into account the Wolf-Rayets with $C/O \leq 1$ (WO) stars, that are produced in a limited mass range (around $60 M_{\odot}$) and only at low metallicity ($Z = 0.010$). In the results presented in \cite{GeorgyC} they end up with a NS or a BH as a remnant depending on the initial progenitor star mass, which is between $15M_{\odot}<M< 32 M_{\odot}$ in the first case, while in the latter $32M_{\odot}<M<60M_{\odot}$. 
However, we here point out that, according to \cite{Kotake2012}, a scenario of $35$ solar masses is preferred.  
Moreover, among the progenitor types they select in the mass ranges between $12M_{\odot}<M<20M_{\odot}$ Red Super Giant (RSG), while for the other cases WC and WNL. \cite{Yoon2006} present grids of stellar evolution models at $Z = 0.004,0.002, 0.001$, and $0.00001$, for rotating magnetized stars in the initial mass range $12M_{\odot} \leq M_{init} \leq 60M_{\odot}$, and with initial equatorial rotation velocities between zero and $80\%$ of the Keplerian value ($0 \leq \frac{v_{init}}{v_K} \leq 0.8$). For $Z=0.001$, the mass range of GRB production is $12 M_{\odot}<M_{init} \leq 30 M_{\odot}$, and $20 M_{\odot}<M_{init} \leq 40 M_{\odot}$. In addition, GRB progenitors have a more massive helium envelope for slower semi-convection, on average. At $Z=0.00001$, on the other hand, the rapidly rotating stars with $M>60 M_{\odot}$ form CO cores of $M >40 M_{\odot}$, which
may be unstable due to the pair instability (cf. \cite{Heger}). The precise CO core mass limit for the pair instability may increase with higher core angular momentum \cite{Glatzel1985}.
 We take these results as a reference for our study to achieve probable dimensionless GW amplitude and the GW emission frequency connected to the GRB. In fact, the mass $M \approx 25M_{\odot}$ is near the transition from NS to BH formation \cite{Mazzali2002}, the exact boundary depending on rotation and mass loss \cite{Tominaga2005}, while we, conservatively, chose the star of $M\approx 10 M_{\odot}$ \cite{Yoon2006}.
\end{itemize}

The paper is organized as follows. In Sec. 2, we discuss the possible GW emission associated to LGRBs. The  goal is to show that the energy emitted in GW by the collapsing star can lead to the determination of the progenitor mass and to a suitable estimate of the fraction  $k << 1$ of the electromagnetic energy $E_{iso}^{GRB}$ that is emitted in GWs, and the energy emitted in GWs by the remnant. In Sec. 3 we discuss the algorithm we used, to obtain the progenitor and remnant parameters from the observed $E_{iso}^{GRB} \pm \sigma_{iso}^{GRB}$. We used the $E_{iso}^{GRB}$ values observed by {\it Swift} \cite{swift}, which span over a very wide redshifts range. Results of calculations are reported in Sec. 4 while Sec. 5 is devoted to discussion and conclusions. In particular we discuss the initial conditions required to obtain collapsars and LGRB.

\section{Gravitational waves emission associated with long Gamma Ray bursts}
%%%%%%%%%%%%%%%%%%%%%%%%%%%%
 The progenitors of LGRBs can be considered as the final step of a stellar collapse and their modeling is rather complicated since cumbersome three-dimensional simulations are required. Many simulations have been performed for intermediate cases of core-collapse of SNae, which should provide GW bursts \cite{Ott}. Gravitational collapse leading to the formation of a NS has long been considered an observable source of GWs. During the core collapse, an initial signal is expected to be emitted due to the changing  of mass distribution. A second part of the GW signal is produced when gravitational collapse is halted by the stiffening of the equation of state above nuclear densities and the core bounces, driving an outward moving shock, with the rapidly rotating proto-neutron star (PNS) oscillating in its axisymmetric normal modes. 
In this paper we show that, at least for our GRB sample, taking into account the GRB distances ($\approx Gpc$, well beyond the horizon of GW antennas), with the Advanced VIRGO and LIGO we will not be able to see GRBs within a certain ranges of frequencies and amplitudes, also according to the energy conversion factor from electromagnetic to GW that we estimate. 

As a first step, we start by evaluating the dimensionless GWs amplitude $h_0$. We remember that, in cases where the GWs emission comes from a  core-collapse process, we cannot predict the exact shape of the emitted GWs signal, but we can have an estimate starting from the energy flux. 
%% FG 22/12 %%
Hereafter we will use the following notation:
\begin{equation}
h(t) = \int_{\infty}^{\infty}{\tilde{h}(f)\,e^{-2\pi i f t}\,df}\,.
\label{eq:hdieffe}
\end{equation}

The energy flux of GWs, defined as the power emitted per unit of surface at a large distance from the source is \cite{Maggiore}:
\begin{equation}
{\cal F}_{\rm GW} = \frac{d^2E^{iso}_{GW}}{dt dS} =\frac{c^3}{16\pi G} \langle (\dot{h}_{+})^2 + (\dot{h}_{\times})^2 \rangle \, ,
\label{eq:flux}
\end{equation}
where <...> indicates a temporal average over a large enough number of periods.
The total energy emitted assuming isotropic emission is then
\begin{equation}
E_{iso}^{\mathrm{GW}} = 4\pi D^2 \int dt \, {\cal F}_{\rm GW} \, ,
\label{eq:EGW}
\end{equation}
where $D$ is the distance to the source. We can write the power $P_{iso}^{\mathrm{GW}}$ emitted in GW by the GRB (the gravitational luminosity), by inserting eq.(\ref{eq:hdieffe}) in eq. (\ref{eq:EGW}) to obtain:
\begin{equation}
 P_{iso}^{\mathrm{GW}} =  \frac{\pi^2c^3}{4G} D^2  \int_{-\infty}^{\infty}{ f\,\tilde{h}_0(f) e^{-2\pi i f t}\,df} \int_{-\infty}^{\infty}{ f'\,\tilde{h}^*_0(f') e^{2\pi i f' t}\,df'}\,,
\label{eq:PGW}
\end{equation}
being 
\begin{equation}
\tilde{h}_0^2(f) =\sqrt{(\tilde{h}_{+})^2 + (\tilde{h}_{\times})^2} .
\end{equation}
If the frequency distribution of the GRB is peaked around a frequency $f_{GW}$, as {\it e.g.} in a sine-Gaussian waveform, eq. (\ref{eq:PGW}) can be written as:
\begin{equation}
 P_{iso}^{\mathrm{GW}} =  \frac{\pi^2c^3}{4G} D^2 2 f_{GW}^2 h_0(t)^2\,.
\label{eq:PGW1}
\end{equation}

Considering that the only observable in the GRB emission is the flux of electromagnetic energy $E_{iso}^{GRB}$ (in a certain band), we make the assumption that the energy emitted in gravitational waves is a fraction $k$ of $E_{iso}^{GRB}$.
So we have:
\begin{equation}
P_{iso}^{\mathrm{GW}} = \frac{k E_{iso}^{GRB}}{T_{90}} , 
\label{T90}
\end{equation}
where $T_{90}$ is the time in which the $90\%$ of $ E_{iso}^{GRB}$ is emitted, and so: 
\begin{equation}
h_{0_{\mathrm{GRB}}}^2(t)\, =\, \frac{k E_{iso}^{GRB}}{T_{90} \pi^2 D^2 f^2_{GW}}{\frac{G}{c^3}}\,.
 \label{hzeroene}
\end{equation}

 The fraction $k$ will also be a tunable parameter useful in order to compare the present and foreseen GW interferometric antenna sensitivities ({\it i.e.} Adv. VIRGO, Adv. LIGO and the forthcoming underground GW Einstein Telescope) to the mean scaled GW amplitudes.
 
Let us consider a progenitor star of mass $M_{PRG}$, radius $R_{PRG}$ and oblateness $\epsilon_{PROG}$, rotating with a frequency $f_{PRG}=0.5f_{PRG_{GW}}$, and a NS as a remnant after the process of GRB emission with parameters $M_{NS}$, $R_{NS}$, $\epsilon_{NS}$ and $f_{NS}$. We have that \cite{Maggiore}:
\begin{eqnarray}
h_{\mathrm{GW}}^{PRG} &=& \frac{4\pi^2G}{c^4}\left(\frac{I_3^{PRG} f^2_{PRG_{GW}}}{D}\epsilon_{PROG} 
\right)\,,
\label{hoBal}\\
h_{\mathrm{GW}}^{NS} &=& \frac{4\pi^2G}{c^4}\left(\frac{I_3^{NS} f^2_{GW_{NS}}}{D}\epsilon_{NS}
 \right)\,.
\label{hoBal1}
\end{eqnarray}
 The momentum of inertia for a spherical rigid object (considering negligible the quadrupolar oblateness $\epsilon $) is given by:
\begin{eqnarray}
I^{NS}_3&=&\frac{2}{5} M_{NS} R_{NS}^2\,, \label{I31}
\\ \nonumber \\
I^{PRG}_3&=&\frac{2}{5} M_{PRG} R_{PRG}^2\,.
\label{IPRG}
\end{eqnarray}
The hypothesis of a rigid body also for the progenitors, can be accepted if one considers that magnetic fields can enhance their stiffness.
Taking into account the total energy balance of the process, we can roughly say that the energy emitted in GW by the GRB is given by the difference of the energies emitted in GRB by the progenitor and the remnant
%\begin{equation}
%E_{iso GW}= E_{PRG}^{GW} - E_{NS}^{GW}
%\end{equation}
resulting in a similar relationship between the squares of GW amplitudes. From eqs. (\ref{hzeroene}), (\ref{hoBal}), (\ref{hoBal1}), (\ref{I31}) and (\ref{IPRG}), we can very roughly write, in terms of the only observables, {\it i.e.} $E^{iso}$ and T90:  
\begin{equation}
E_{iso}^{GRB}= \frac{64 G \pi^6 T_{90} f_{GW}^2 \left(M_{PRG}^2 R_{PRG}^4 f_{PRG}^4 \epsilon_{PRG}^2-M_{NS}^2 R_{NS}^4 f_{NS}^4 \epsilon_{NS}^2\right)}{25 c^5 k}\,,
\label{eq:balance}
\end{equation}
so we get a function useful to solve our problem, that is 
\begin{equation}
E_{\mathrm{GW}}^{iso}=\, f( k ,\epsilon_{NS}, M_{NS} ,R_{NS},f_{{NS}} , \epsilon_{PRG}, M_{PRG}, R_{PRG},f_{PRG}, f_{GRB})\,.
\label{MISO}
\end{equation}
To evaluate the physical parameters involved in the GRB process to the remnant phase, we should estimate the $N=10$ parameters in eqs. (\ref{eq:balance}) having only one observed data {\it i.e.} $E_{\mathrm{iso}}^{GRB}$. However the assumption we make for the progenitor models are very well posed in literature so that we can set very reasonable ranges for $9$ parameters out of $10$, therefore the only key parameter we infer is the scale factor $k$.\par
 In the next section we will put the bases to solve this problem using a key idea and a suitable algorithm. 
 
 %%%%%%%%%%%%%%		
\section{Genetic controlled random search algorithm for GRB progenitor masses estimate} 
\begin{figure}
\center
\begin{tabular}{|l|l|}
\hline
\\
\includegraphics[width=8 cm, height=8 cm]{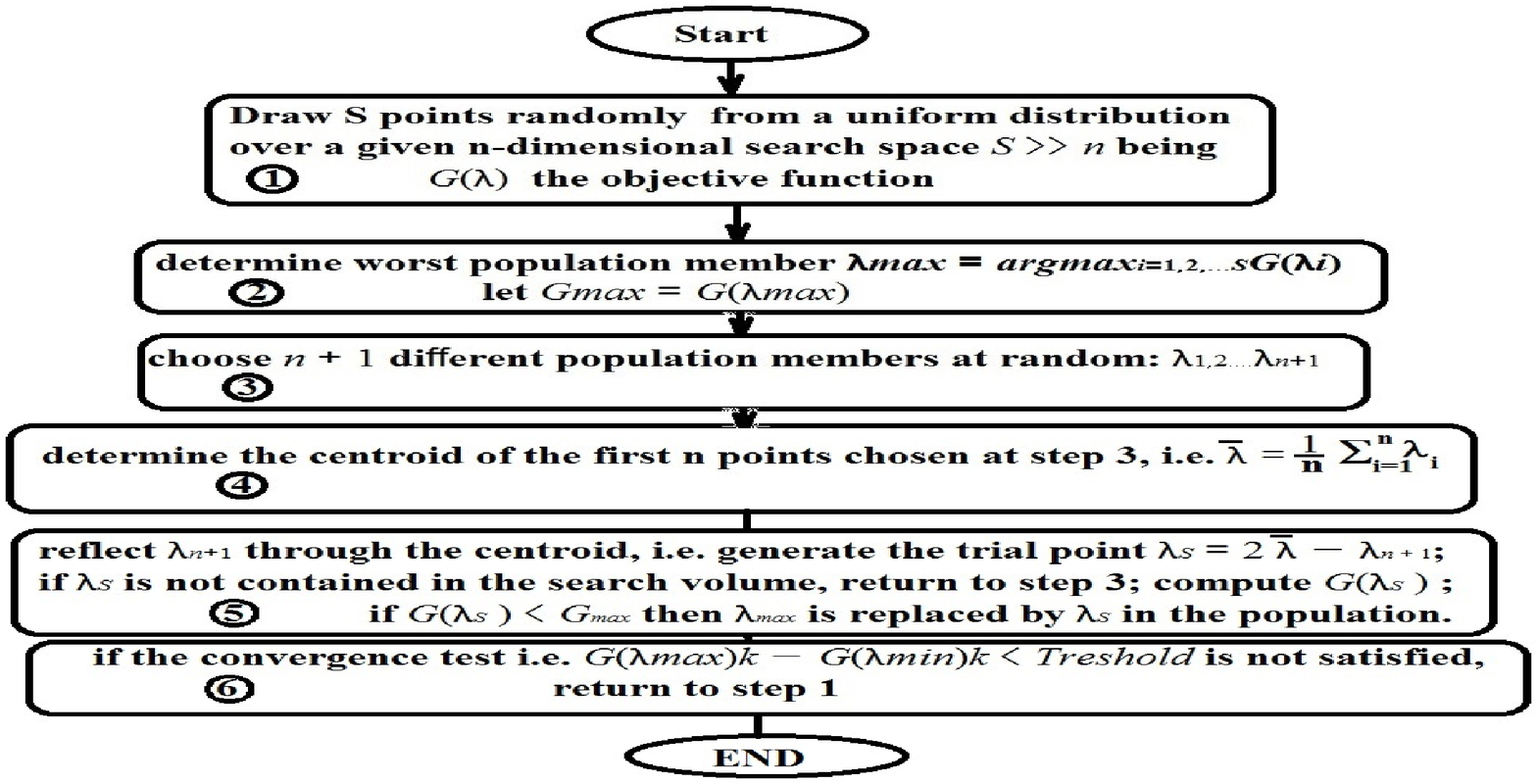}
\includegraphics[width=8 cm, height=8 cm]{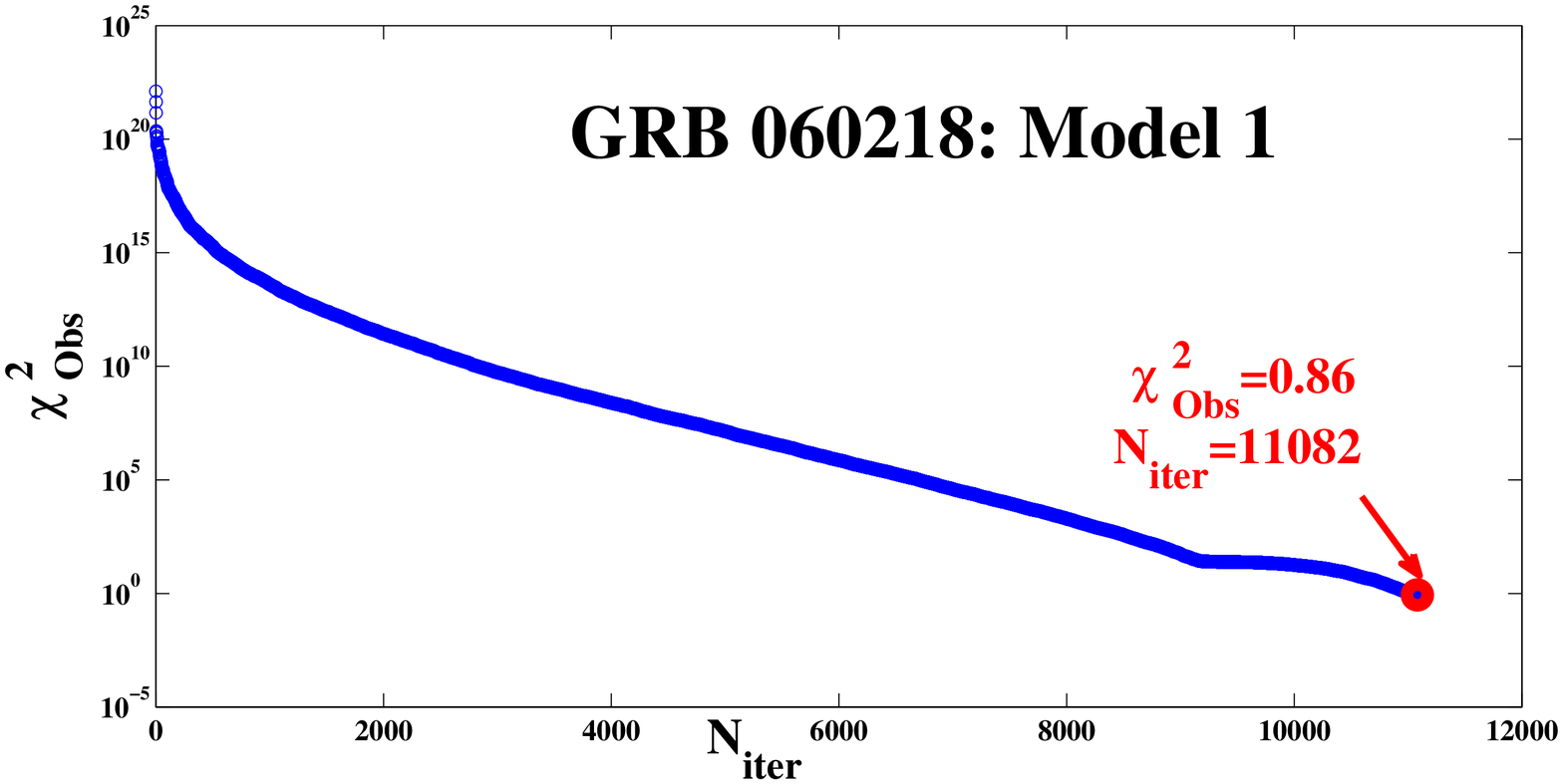}
\\
\hline
\includegraphics[width=8 cm, height=8 cm]{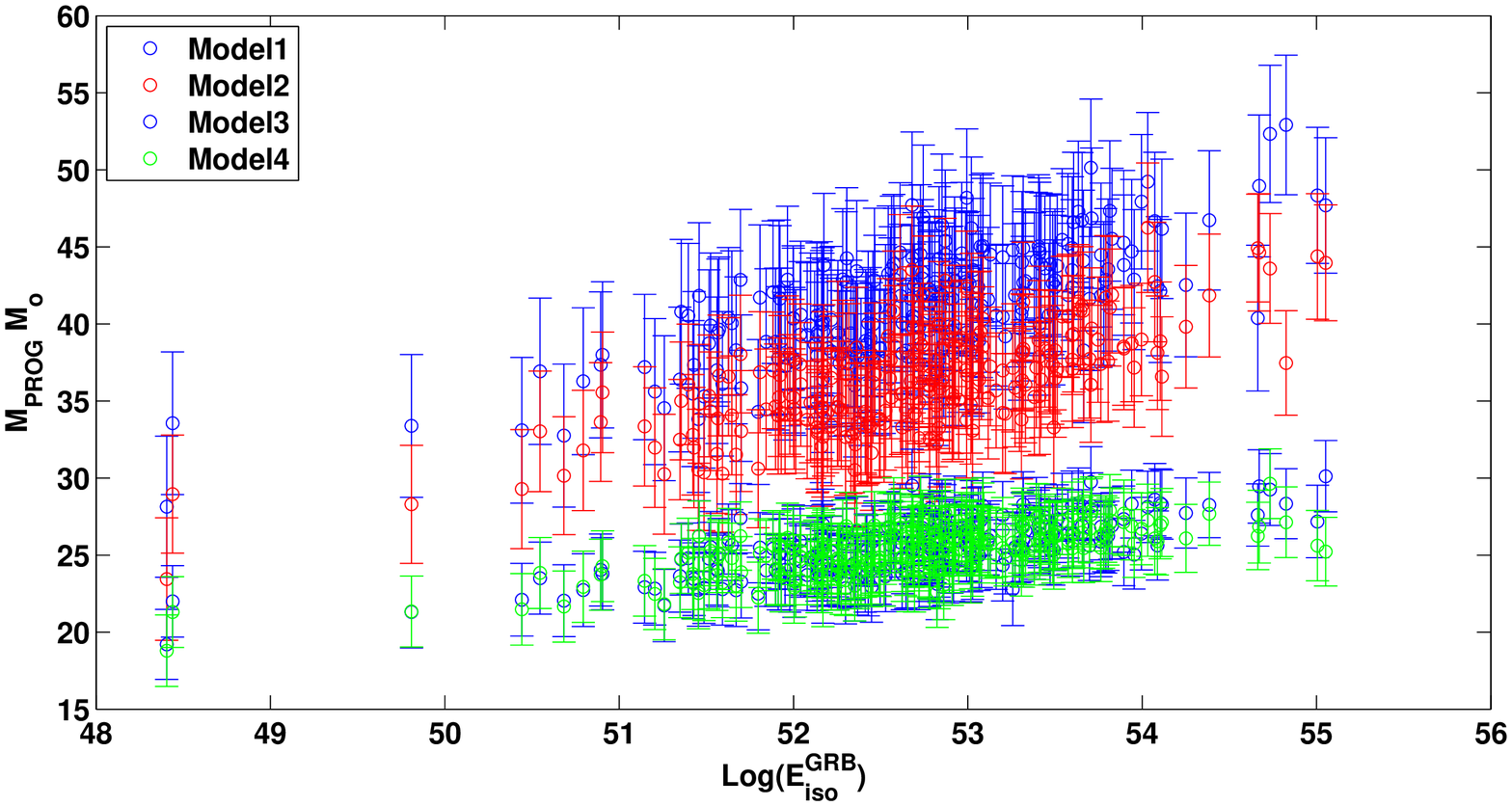}
\includegraphics[width=8 cm, height=8 cm]{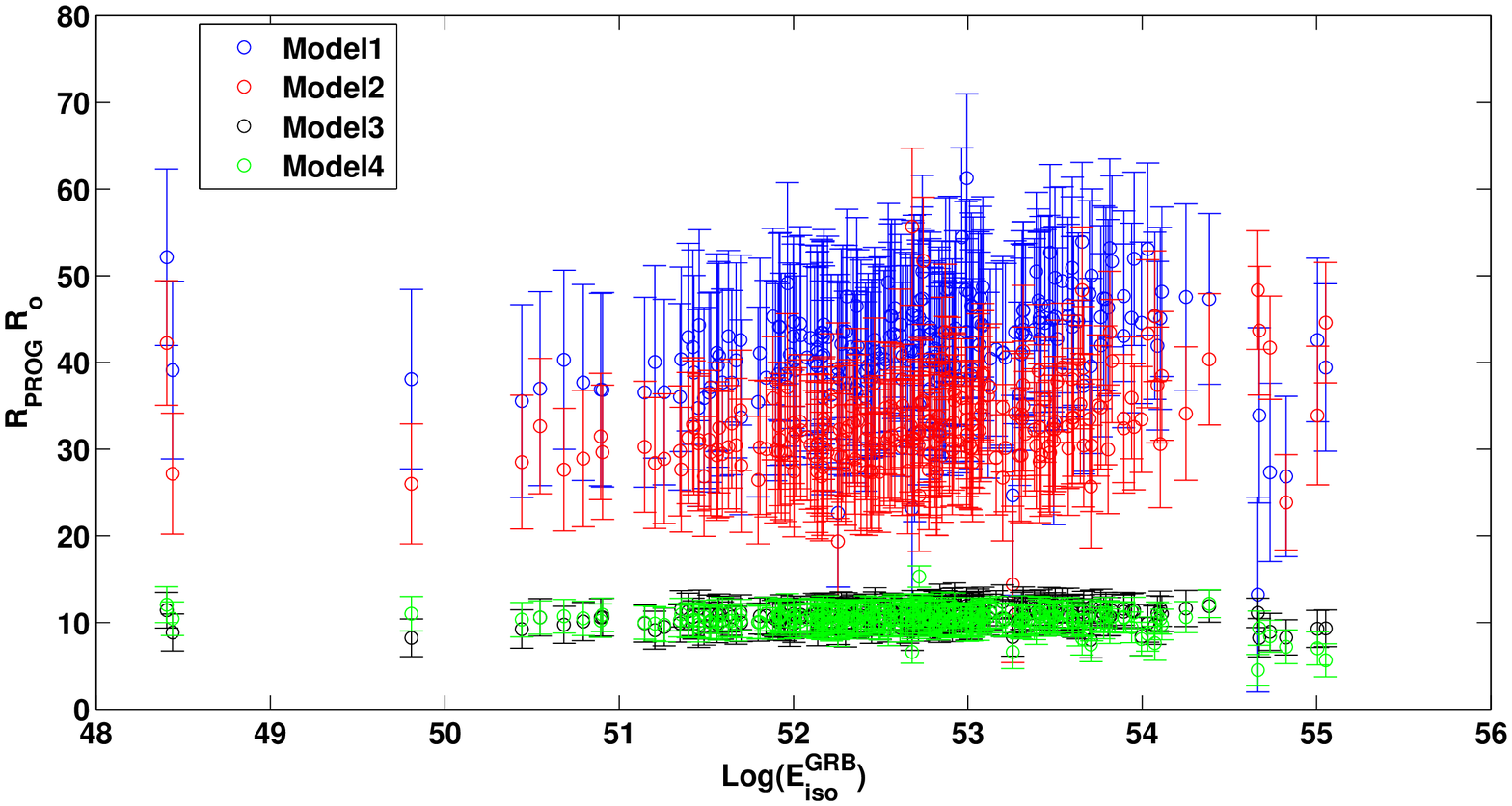}
\\
\hline
\end{tabular}
\caption{ On the top panel we show the flux diagram of the algorithm,  and  an example of convergence at confidence level of $5 \%$  $\chi ^{2}_{fit}\leq\chi ^{2}_{Obs} \approx 0.86$ (red circle) for the test case of GRB$060218$.  On the bottom the four region of possible solutions for the progenitor masses and radii are shown according to the used parameters range.}
\label{Exampl_Conv}
\end{figure}

Our aim, as stated above, is to find the best progenitor mass starting from the observed $E_{iso}^{GRB}$.
Given eq. (\ref{eq:balance}) we see that, from the observed $E_{iso}^{GRB}$ we must infer ten parameters:
\begin{enumerate}
		\item The factor $k$ to evaluate the scaled $E_{iso}^{GW}$  from the observed $E_{iso}^{GRB}$;
		\item The emission frequency $f_{GW}^{GRB}$; 
		\item The estimate of the progenitor quadrupolar oblateness. According to interferometric measures and stellar evolution grids, it can be put in the interval $\displaystyle{0.01 \div \frac{2}{3}}$ being the upper limit obtained from the star Roche lobe limit, at the critical velocity  \cite{MeynetG}.
		\item The mass of the progenitor. According to Grids of stellar models with and without rotation \cite{EkstromS,GeorgyC} we  assume the progenitor mass in the interval $ 10 \div 60 M_{\odot}$.		
		\item The frequency of GW emission of the progenitor. According to the Grids of stellar models \cite{EkstromS,GeorgyC} we assumed it in the range: $10^{-10}  \div 10^{-6}Hz$  inferred from the critical velocity according to the Roche model \cite{MeynetG}.
		\item The estimates of the remnant quadrupolar oblateness. According to the current NS model and the corresponding Equation of State (EoS) it can be set in the interval $10^{-4}\div 10^{-8}$  
		\cite{VO}.
		\item  The mass of the remnant. According to different EoS models the mass of the remnant could be in the range $1.0\div 3 M_{\odot}$ \cite{VO,Kalogera} or for a stellar BH in the range $1.0\div 10 M_{\odot}$ \cite{EkstromS} . 
		\item The radius of the progenitor. According to the Grids of stellar models \cite{EkstromS,GeorgyC} and references therein) we can assume stellar radii of the progenitor in the interval $0.9 \div 140 R_{\odot}$ from stellar model of mass ranges $ 10 \div 60 M_{\odot}$.
		\item  The radius of the remnant. According to different EoS models, the radius of the remnant could be in the range $5 Km \leq R \leq 15 Km $, \cite{VO} or for a stellar BH in the range $5 Km \leq R \leq 30 Km $.
		\item  The frequency of GW emission of the remnant. From the known frequencies of electromagnetic energy emission by pulsars we can consider GW frequencies $f_{NS}$  in the range $0.10 \div 1000 Hz$ 
		\end{enumerate}
To solve the problem we use an algorithm that is capable to find the global minimum of a multivariable function. We used a simplified simplex algorithm: the controlled random search (CRS) algorithm and, 
 to improve its performances, we used a genetic modification of the search procedure making the software more resilient to local minima (\cite{Price,Milano1997,Milano2002,M_Bresco} ). In Fig. \ref{Exampl_Conv} we show the flux diagram of the algorithm.

First of all we must obtain an objective function to minimize, using the reduced $\chi_{obs}^2$ with $N_{GRB}-N$ degrees of freedom, where $N=10$ is the number of free parameters:
\begin{equation}
F_{Ob}=\, \chi_{obs}^2=\frac{1}{N_{GRB}-N} \sum^{N_{GRB}}_{i=1} \frac{\big (E_{{iso}_{i_{Obs}}}-E_{{iso}_{i_{C}}})^2}{\sigma_{{iso}_{i}}^{2}}\,,
\label{OBJ}
\end{equation}
where $E_{{iso}_{[C, Obs]}}$ is given by Eq. (\ref{eq:balance}).
It worths noticing that we have only one observed datum {\it  i.e.} $E_{iso_{Obs}} \pm \sigma_{E_{iso}}$ per GRB, in other words one equation and ten unknown parameters to estimate. We solve the problem by a Montecarlo procedure, for each observed GRB, we generate a large number $N_{GRB}$ of $E_{iso}$ values extracted randomly from a Gaussian distribution with mean value $E_{{iso}_{Obs}} $ and standard deviation $\sigma_{E_{iso}}$ given by the observed measurement error:
\begin {equation}
E_{iso_{gen}}=E_{iso_{Obs}} + \sigma_{Eiso} randn(N_{GRB})\,,
\label{eisosim}
\end {equation}
then we evaluate the reduced $\chi^{2}_{fit}\leq \chi ^{2}_{Obs}$ with $N_{GRB}-N $ degrees of freedom at $5\%$ of confidence level ($\chi ^{2}_{fit} \leq 0.86$).
%i.e. $\chi ^{2}_{fit} \leq 1.3$.
Finally we evaluate the mean $E_{iso_{gen_M}}$ of $E_{iso_{gen}}$ to check the relative difference between the Observed (O) and calculated (C) values:   
$ O-C= \frac{E_{iso_{Obs}} - E_{iso_{gen_M}}}{E_{iso_{Obs}}}$ of the four regions of possible solutions, hereafter indicated as model $1...4$,  with parameters in the interval shown in Tab.\ref{tableModel}. 
\begin{table*}
\centering
\caption{Model parameters ranges used according stellar evolution model of \cite{MeynetG}}
\begin{tabular}{|l|l|l|l|l|l|l|l|l|}
\hline

 $ Name$	&	$M_{PROG}\, (M_{\odot})$	&	$ R_{PROG}\, (R_{\odot})$	& $M_{Remn} \,(M_{\odot})$	&	$R_{Remn}\, (km)$	& $f_{Remn} \,(Hz)$ \\
\hline

Model 1 &	$10\div 60$	&	$0.9\div 140$	&	$1\div 3$	&	$6\div 15	$ & $0.1\div 1000$\\
\hline
Model 2 &	$10\div 60$	&	$0.9\div 140$	&	$1\div 10$	&	$6\div 30	$ & $0.1\div 1000$\\
\hline
Model 3 &	$10\div 35$	&	$0.9\div 25$	&	$1\div 3$	&	$6\div 15	$ & $0.1\div 1000$\\
\hline
Model 4 &	$10\div 35$	&	$0.9\div 25$	&	$1\div 10$	&	$6\div 30	$ & $0.1\div 1000$\\
\hline
\end{tabular}
\label{tableModel}
\end{table*}
 The parameters common to the different models of Tab.\ref{tableModel}, to initialize the domain are shown in Tab. \ref{tableComPar}.
%%%%%%%%%%%%%%%%%%%%%%%%%%%%%%%%%%%%%
\begin{table*}
\centering
\caption{Parameters common to the different model of Tab. \ref{tableModel} useful to initialize the domain of possible solutions.}
\begin{tabular}{|l|l|l|l|l|l|}
\hline
Model& $k$	&	$ f_{GRB}^{GW}\, (Hz)$	&$\omega_{PROG}\,	(rad/s)$& $\epsilon_{PROG} $&	$\epsilon_{Remn} $ \\
\hline
Model C&$10^{-9}\div 10^{-4}$ & $50\div 800$ &$10^{-4}\div 10^{-10}$& $10^{-2}\div {\frac{2}{3}}$ & $10^{-8}\div 10^{-4}$\\
\hline
\end{tabular}
\label{tableComPar}
\end{table*}

Using these ranges of parameters we analyzed a sample of $237$ GRB and the results are treated in the next section.
In Fig. \ref{Exampl_Conv} we show the flux diagram of the algorithm, and an example of convergence curve using the value of the $\chi ^{2}_{Obs}$ as a function of the number of iterations. The red circle, represents $\chi ^{2}_{fit}\leq\chi ^{2}_{Obs} \approx 0.86$ giving a confidence level of $5 \%$ for the test case of GRB $060218$ that we will analyze later.
  
%%%%%%%%%%%%%%%%%%%%%%%%%%%%%%%%%%%%%%%%%%
\section{Gamma ray bursts sample and results on gravitational wave emission}
%%%%%%%%
%tre
%%%%%%
We selected a sample of $237$ long GRBs with known redshifts that we then analyzed, detected by the {\it Swift} satellite \cite{swift} from January 2005 to May 2014. This sample is an extended sample presented in \cite{Dainottia,Dainottib, Dainotti2015} where GRBs with plateaus are presented.
We use the GRB Coordinates Network \cite{GCN} for redshifts and exclude GRBs with non-spectroscopic redshifts. In Tab. \ref{Sample} the sample GRB data range, that we analyzed, is shown. 
\begin{table*}
\centering
\caption{Sample GRB data range used in this work}
\begin{tabular}{|l|l|l|l|l|l|}
\hline
$E_{iso}^{GRB} \, (erg)$	& $\sigma_{E_{iso}^{GRB}}\,(erg)$ &	$T90\,(s)$	& $\sigma_{T90}\,(s)$ &	redshift $z$&	$D \times10^{26}\,(cm)$\\
\hline
$ 2.5\times10^{48}\div1.1\times10^{55} $&	$ 3.9\times10^{47}\div1.7\times10^{54}$& $2.2\div 844 $&	$0.003\div3$&$0.014\div8.2$&	$1.8\div 2500$
\\
\hline
\end{tabular}
\label{Sample}
\end{table*}

\begin{figure*}
\center
\begin{tabular}{|l|l|}
\hline
\\
\includegraphics[width=8 cm, height=6 cm]{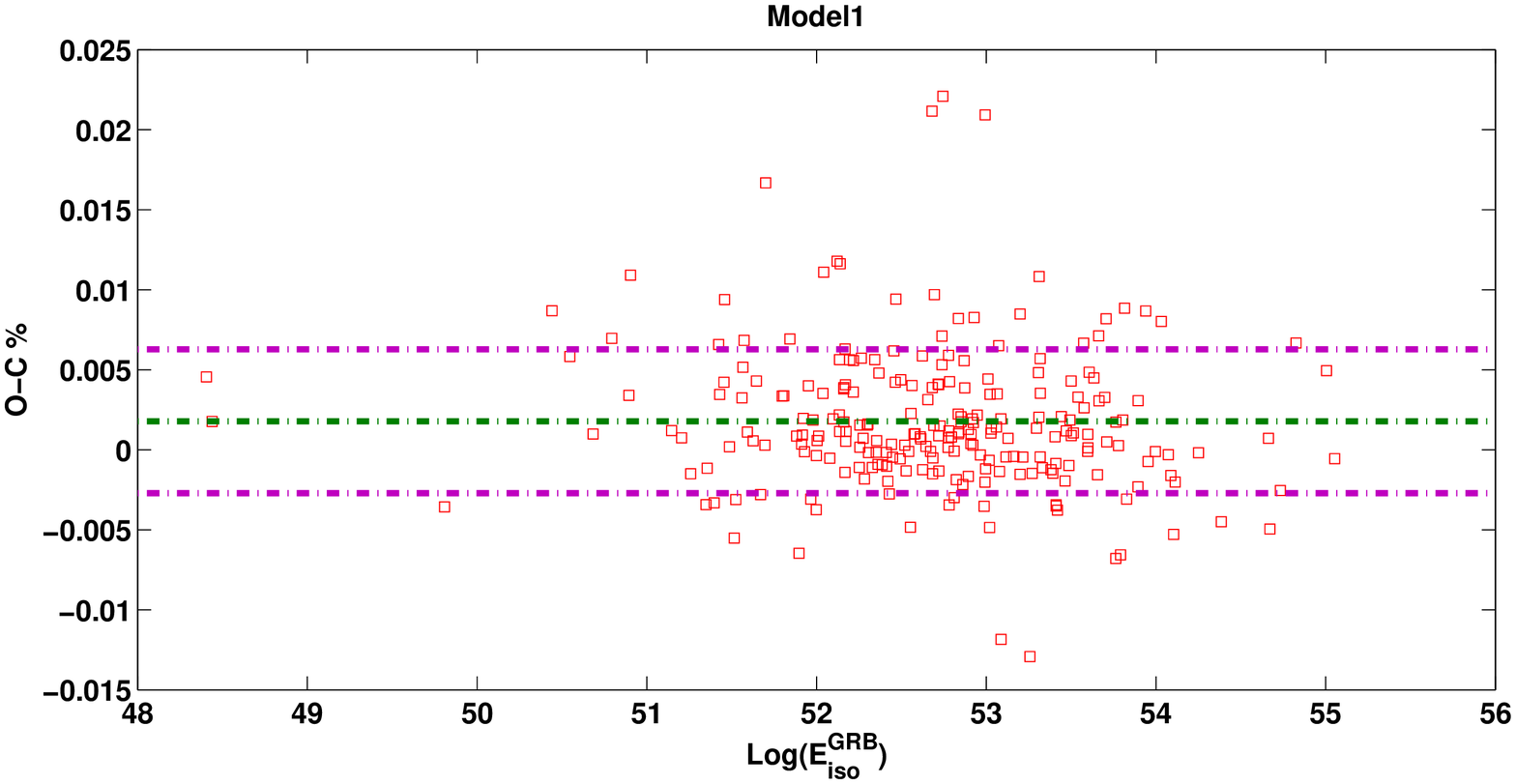}
\includegraphics[width=8 cm, height=6 cm]{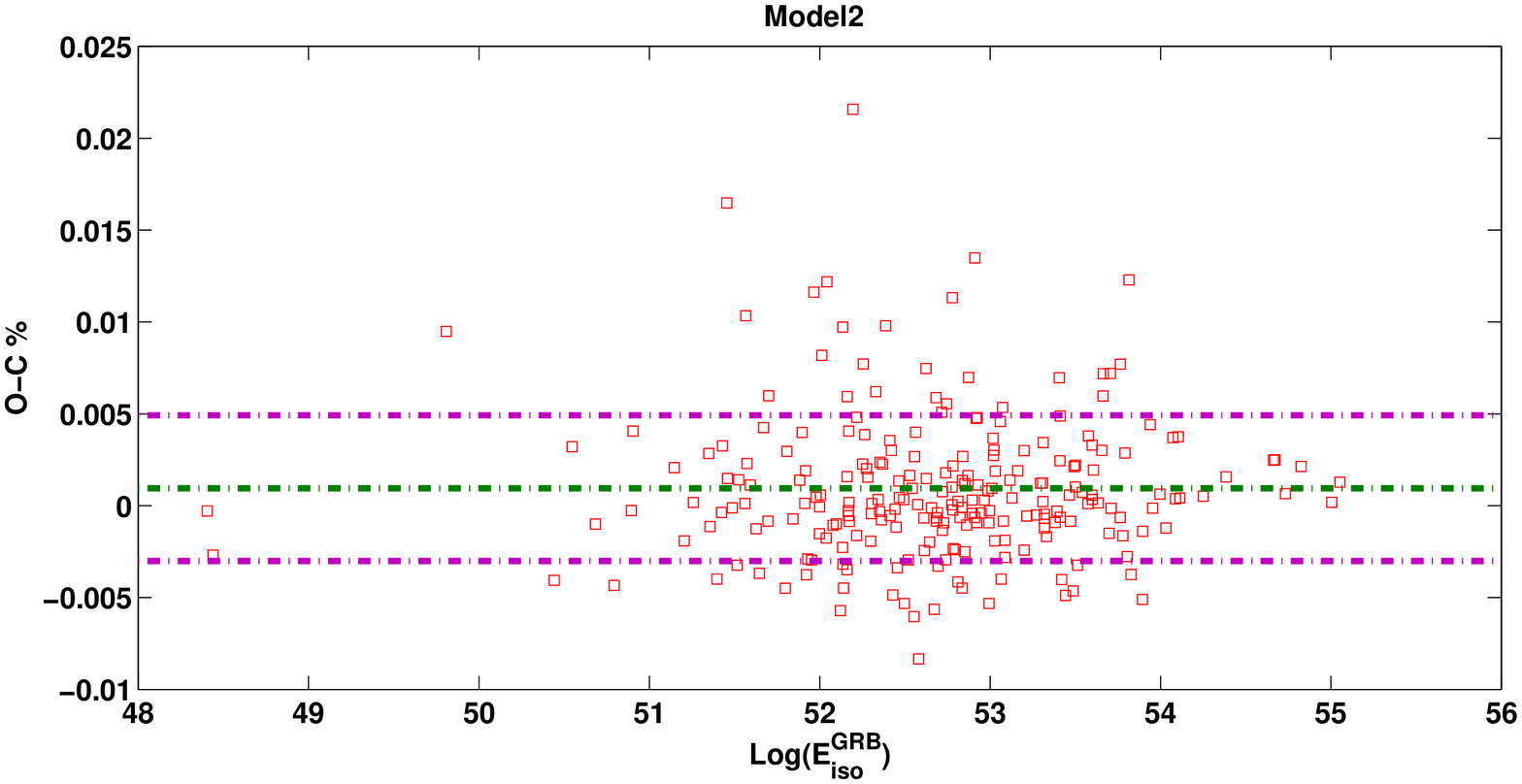}\\
\hline
\includegraphics[width=8 cm, height=6 cm]{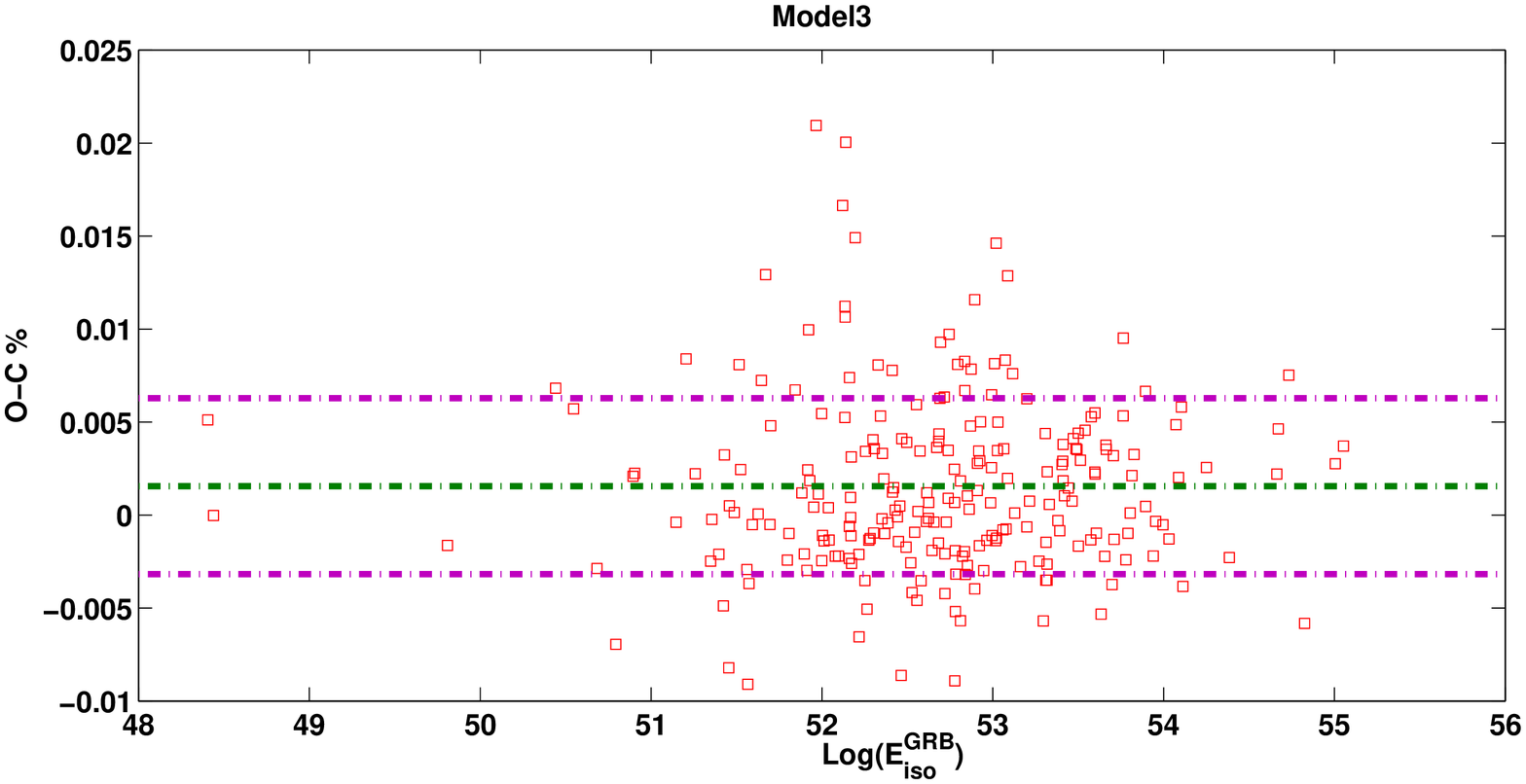}
\includegraphics[width=8 cm, height=6 cm]{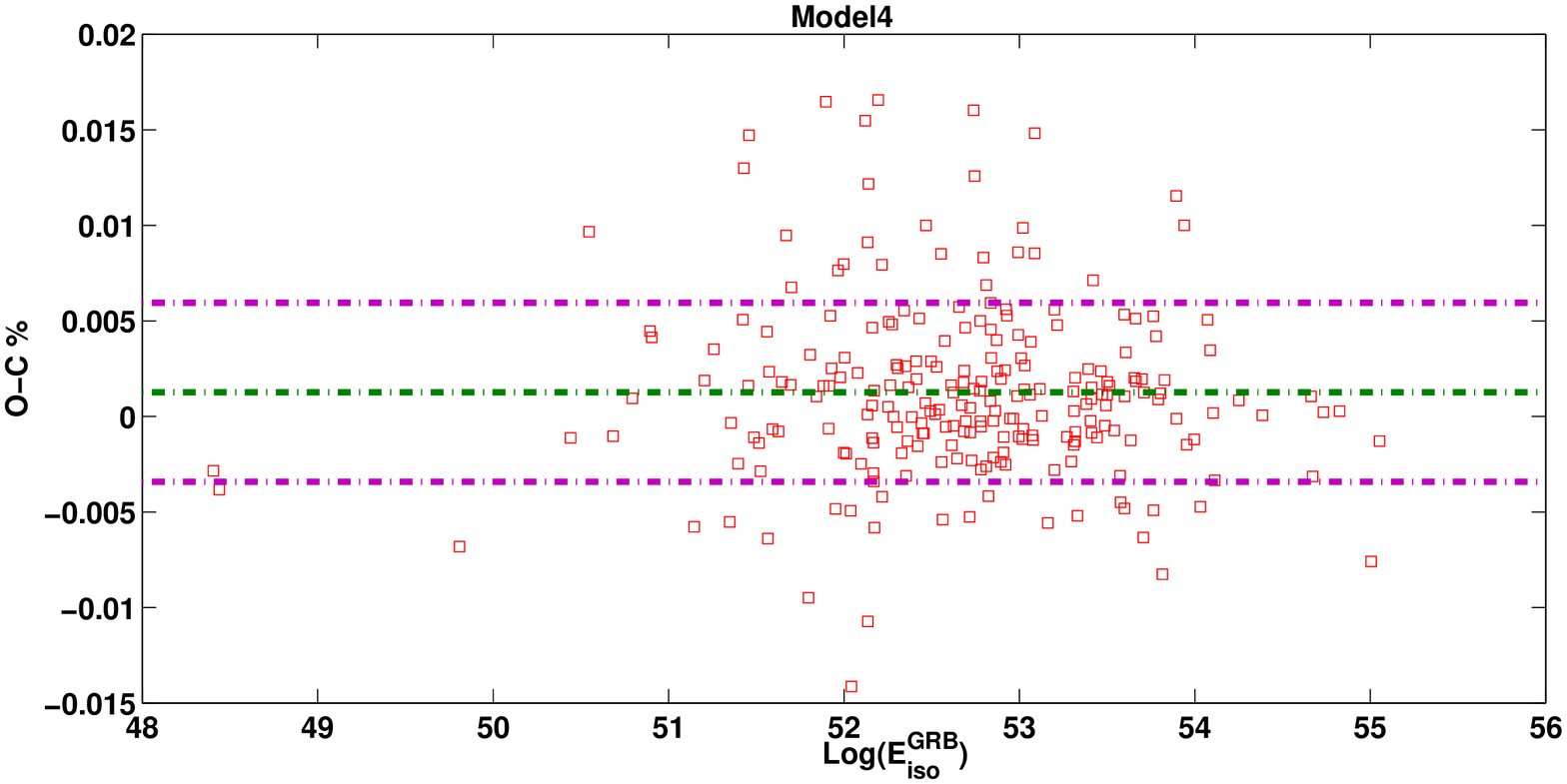}\\
\hline
\end{tabular}
\caption{Residuals $\displaystyle{ O-C= \frac{E_{iso_{Obs}} -E_{iso_{C}}}{E_{iso_{Obs}}}}$  of the fitting against  $Log(E_{iso_{Obs}})$ for the four model with parameters interval shown in Tab.\ref{tableModel}. It is evident that the agreement is rather good. }
\label{OMC}
\end {figure*}

Considering the above mentioned sample of long GRBs and  according to the hypotheses we made, it is possible to infer the physical parameters of the GRB progenitor with their errors. The mean results of our analysis are shown in Tab. \ref{tableResult} and Tab. \ref{tableResultCommon}  whilst in Fig. \ref{OMC} the residuals  $O-C \%$ of the fitting against the $E_{iso_{Obs}}$  are shown in log scale for the four models with parameters interval shown in Tab.\ref{tableModel} and Tab. \ref{tableComPar}. It is evident that the agreement is fairly good.

%%%%%%%%%%%%%%%%%%%%%%%%%%%%%%%%%%%%%%%%%%%%%%%%%%%%%%%%%%%%%%%%%%%%%%%%%%%%%%%%%%%%%%%%%%%%%%%%%%%%%%%%%%%%%%%%%%%%%%%%%%%%%%%%%%%%%%%%%%%%%%%%%%%%%%%%%%%%%%%%%%%%%%%%%%%%%%%%%%%%%%%%%%%%%%%%%%%%%%%%%%%%%%%%%%%%%%%%%%%%%%%%%%%
\begin{table*}
\centering
\caption{Mean results for model parameters obtained from genetic-Price algorithm}
\begin{tabular}{|l|l|l|l|l|l|l|l|l|}
\hline

 $ Model$	&	$M_{PROG}\,(M_{\odot})$	&	$ R_{PROG} \,(R_{\odot})$	& $M_{Remn} \,(M_{\odot})$	&	$R_{Remn}\, (km)$	& $f_{Remn} \,(Hz)$ \\
\hline
 Model 1&$41\pm 5$	&	$50\pm 12 $	&	$2.3\pm 0.7 $	&	$10\pm 2 $ & $504\pm260 $\\
\hline
 Model 2 &$34\pm 3 $	&	$36\pm 11 $	&	$5\pm 2 $	&	$16\pm 6 $ & $782\pm 300$\\
\hline
Model 3&$ 25\pm 4 $	&	$11\pm 4 $	&	$2.0\pm 0.5 $	&	$9\pm 3$ & $495\pm 200$\\
\hline
Model 4&$25\pm 4 $	&	$10\pm 4 $	&	$5\pm 2 $	&	$15\pm 6 $ & $493\pm 230 $\\
\hline
\end{tabular}
\label{tableResult}
\end{table*}
  The mean results of parameters common to the different models of Tab.\ref{tableResult} are shown in Tab. \ref{tableResultCommon}.
%%%%%%%%%%%%%%%%%%%%%%%%%%%%%%%%%%%%%
\begin{table*}
\centering
\caption{Mean result Parameters common to the different models of Tab. \ref{tableResult} }
\begin{tabular}{|l|l|l|l|l|l|}
\hline
Model& $k$	&	$ f_{GRB}^{GW}\, (Hz)$	&$\omega_{PROG}\,	(rad/s)$& $\epsilon_{PROG} $&	$\epsilon_{Remn} $ \\
\hline
Model 1 &$2\pm1\times10^{-6}$	&	$602\pm 255 $	&$3\pm2 \times10^{-5}$	&	$0.15\pm0.06$	&	$3\pm2\times10^{-6} $ \\
\hline
Model 2&$3\pm 2\times10^{-6}$	&	$593\pm 264$	&$5\pm2 \times10^{-5}$	&	$0.21\pm0.09$	&	$2\pm1\times10^{-6} $ \\
\hline
Model 3&$4\pm2\times10^{-6}$	&	$628\pm270$	&$8\pm5 \times10^{-5}$	&	$0.23\pm0.09$	&	$1\pm0.8\times10^{-5} $ \\
\hline
Model 4&$4\pm2\times10^{-6}$	&	$618\pm 278$	&$9\pm6 \times10^{-5}$	&	$0.22\pm0.08$	&	$5\pm3\times10^{-6} $ \\
\hline
\end{tabular}
\label{tableResultCommon}
\end{table*}

%%%%%%%%%%%%%%%%%%%%%%%%%%%%%%%%%%%%%%%%%%%%%%%%%%%%%%%%%%%%%%%%%%%%%%%%%%%%%%%%%%%%%%%%%%%%%%%%%%%%%%%%%%%%%%%%%%%%%%%%%%%%%%%%%%%%%%%%%%%%%%%%%%%%%%%%%%%%%%%%%%%%%%%%%%%%%%%%%%%%%%%%%%%%%%%%%%%%%%%%%%%%%%%%%%%%%%%%%%%%%%%%%%%
 
\begin{figure*}
\center
%\hline
%\\
\includegraphics[width=10 cm, height=10 cm]{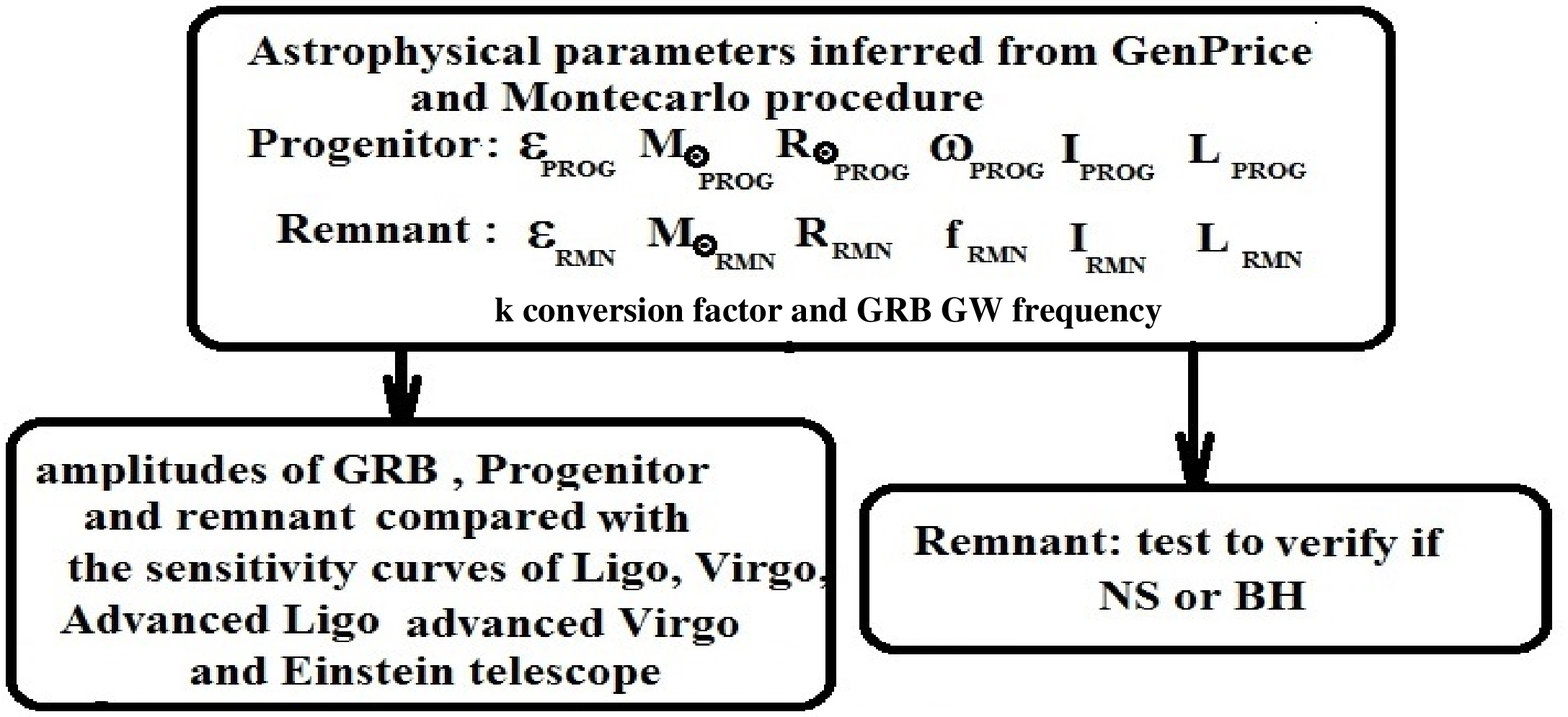} \\
%\hline
\caption{ Astrophysical parameters determined with our hybrid Montecarlo procedure and scheme followed to analyze the results. }
\label{schemaS}
\end{figure*}

 So we have the astrophysical parameters of the possible progenitors and remnants and the GW emission  frequency of GRB.  The scheme of our procedure is shown in Fig. \ref{schemaS}

%%%%%%%%%%%%%%%%%%%%%%%%%%%%%%%%%%%%%%%%%%%%%%%%%%%%%%%%%%%%%%%%%%%%%%%%%%%%%%%%%	

\begin{figure}
\center
\begin{tabular}{|l|l|}
\hline
\\
\includegraphics[width=7 cm, height=7 cm]{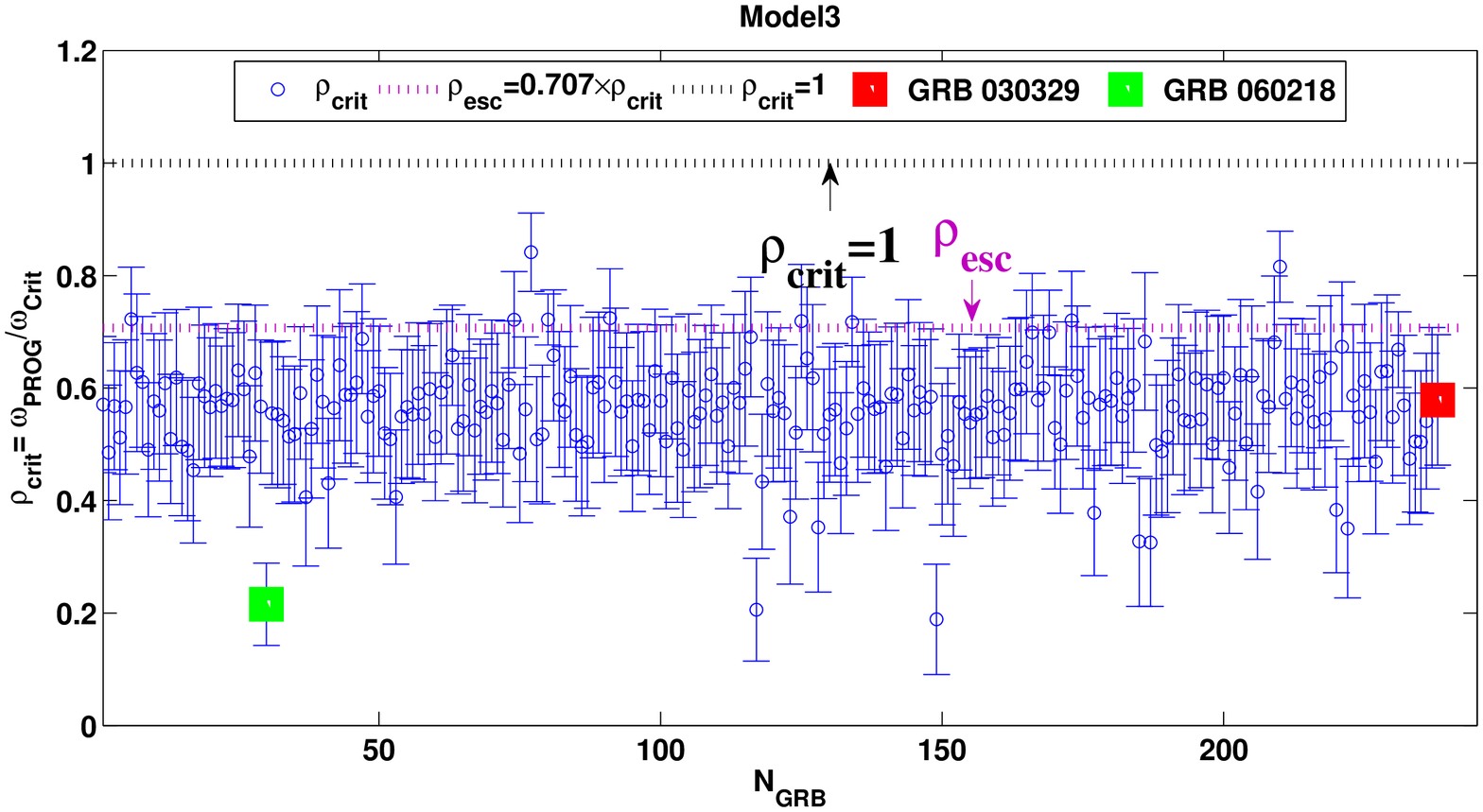}
\includegraphics[width=7 cm, height=7 cm]{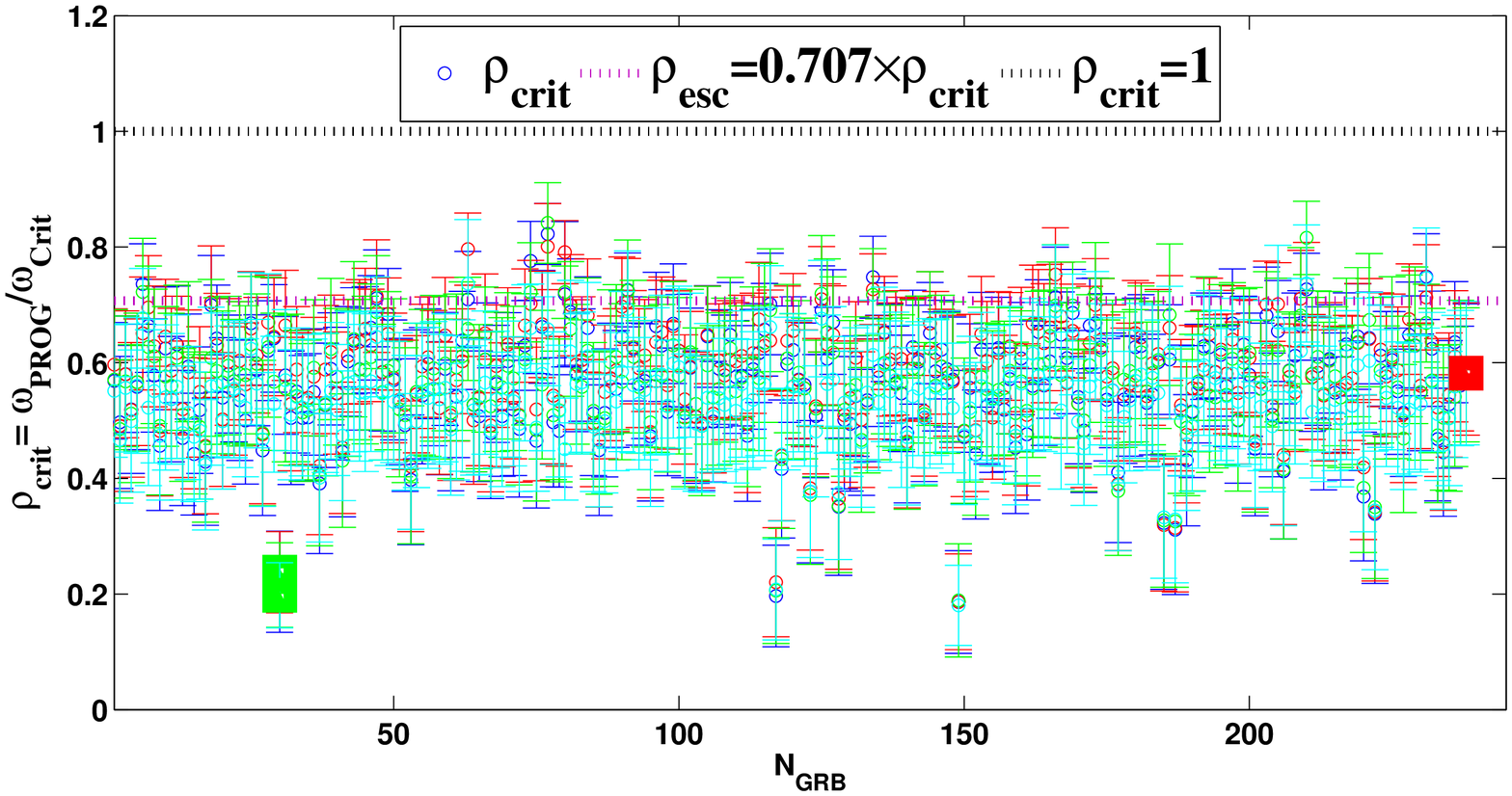}
\\
\hline
\end{tabular}
\caption{ The ratios $\rho_{crit}=\frac{\omega_{PRG}}{\sqrt{\frac{G M}{R^3}}}$ are shown to ascertain whether the progenitor star is rotating at an angular velocity greater than the keplerian velocity. The results are shown, for the model 3 on the left and the global one on the right with the error bar on $\rho_{crit}$. The red  and green squares are the $\rho's$ of GRB$030329$ and GRB$060218$ respectively. The black line marks $\rho_{crit}=1$ whilst the magenta line is the escape ratio $\rho_{escape}=\frac{\omega_{PRG}}{\sqrt{2}\rho_{crit}}$ As it is possible to see the dynamical state of the progenitors is practically the same for all the models.}
\label{OM_Crit}
\end{figure}

The results obtained so far can be analysed in order:
\begin{itemize}
\item to discriminate between stability or instability condition for the progenitor, for the region of possible solutions spanned by genetic-price algorithm;
\item to discriminate between NS or BH solution for the remnant;
\item to discriminate among the possible region of solutions obtained according to the models in analysis.
\end{itemize}
\paragraph{\bf - Stability or instability condition for the progenitor.}
 To discriminate between stability or instability condition for the progenitor, within the region of possible solutions spanned by genetic-price algorithm we used the ratio between the progenitor star angular velocity and the Keplerian velocity (assumed as critical velocity) $\displaystyle{\omega_{PRG}\leq\omega_{crit}=\sqrt{\frac{G M}{R^3}}}$. We compared the unity ratio {\it i.e.} $\displaystyle{\rho_{crit}=1}$ and the escape ratio $\displaystyle{\rho_{esc}=\frac{\rho_{crit}}{\sqrt{2}}}$ to the ratio between the actual angular velocity of the progenitor and the corresponding critical one.
In Fig.\ref{OM_Crit}, the ratios $\displaystyle{\rho_{crit}=\frac{\omega_{PRG}}{\sqrt{\frac{G M}{R^3}}}}$ are shown to ascertain whether the progenitor star is rotating at an angular velocity greater than the Keplerian velocity. The results are shown, for the model $3$ on the left and the global ones on the right with the error bar on $\rho_{crit}$. The black line represents $\rho_{crit}=1$ whilst the magenta line is the escape ratio $\displaystyle{\rho_{escape}=\frac{\omega_{PRG}}{\sqrt{2}\rho_{crit}}}$. As it is possible to see we have values below the critical and near the escape ratio. It is evident that the dynamical state of the progenitors is essentially the same for all the models. The red and the green squares are the $\rho$ of GRB$030329$ and GRB$060218$ respectively, taken as test cases. 

\begin{table*}
\centering
\caption{Sample GRB data for test on GRB$030329$ and GRB$060218$}
\begin{tabular}{|l|l|l|l|l|l|l|}
\hline
Name&$E_{iso}^{GRB} \, (erg)$	& $\sigma_{E_{iso}^{GRB}}\,(erg)$ &	$T90\,s$	& $\sigma_{T90}\,(s)$ &	redshift $z$&	{\it lum. dist.} $D \,(cm)$\\
\hline
$GRB030329$&$ 1.80\times10^{52}$ & $ 7.00\times10^{50}$& $41.02$&	$0.02$&$ 0.1685$&	$1.81\times10^{27}$
\\
\hline
$GRB060218$&$2.54\times10^{48}$ &	$5.22\times10^{47}$ &	128 &$0.2$&	0.0331&$4.42\times10^{26}$
\\
\hline
\end{tabular}
\label{GRBT}
\end{table*}

%%%%%%%%%%%%%%%%%%%%%%%%%%%%%%%%%%%%%%%%%%%%SI/

\begin{figure}
\center
%\begin{tabular}{|l||}
%\hline
%\\
\includegraphics[width=15 cm, height=10 cm]{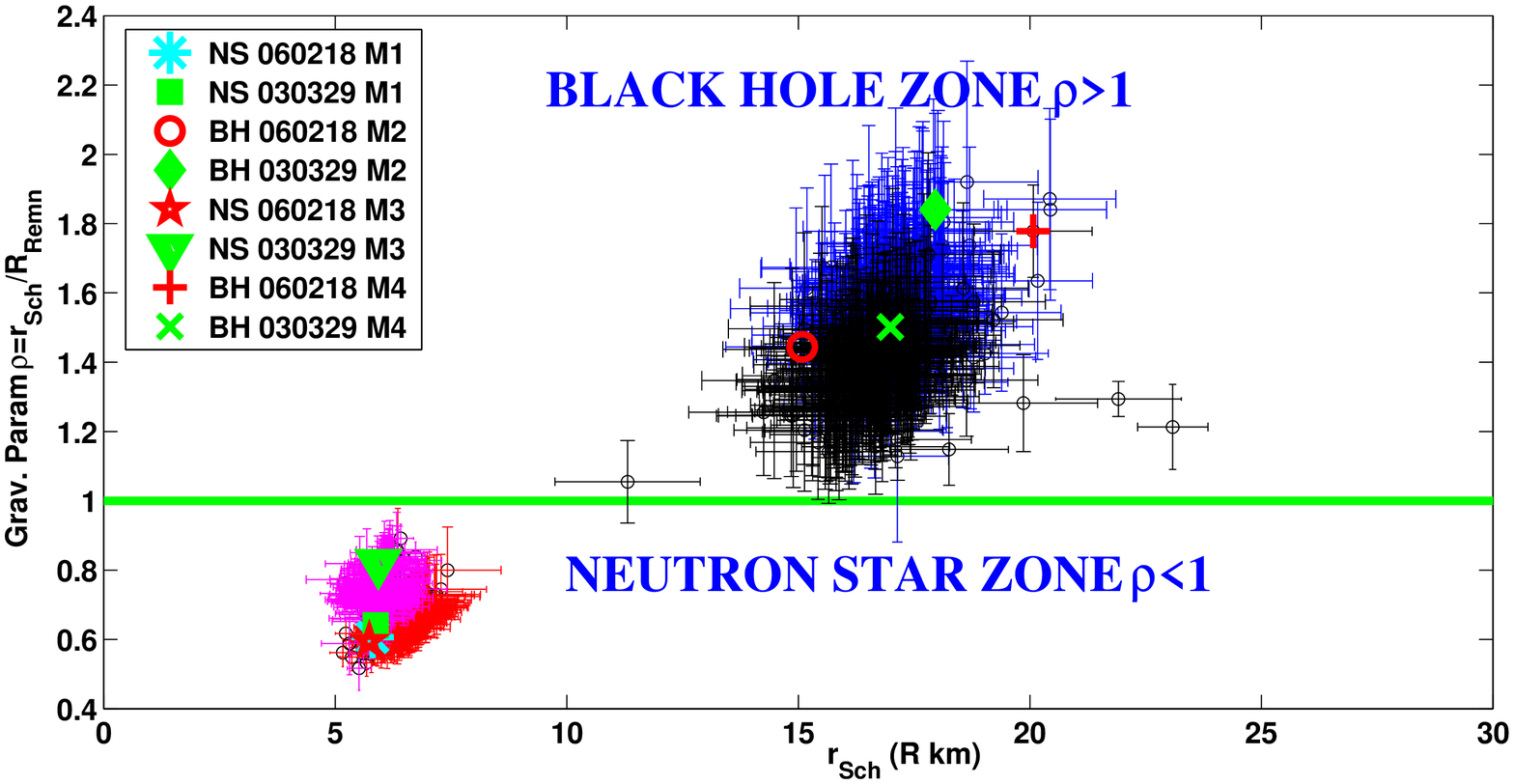}
%\\
%\end{tabular}
\caption{We show here the gravitational parameter $\displaystyle{\rho_{Sch} =\frac{R_{Sch}}{R_{Remn}}}$ being $\displaystyle{R_{Sch}=\frac{2 G M}{c^2}}$ the Schwarzschild radius $R_{Remn}$ and the remnant radius in order to discriminate whether the remnant is a BH or a newborn millisecond pulsar. The black  and the blue circles are the results of Model $2$ and Model $4$ (BH zone) whilst the magenta and red circles are the results of Model $1$ and Model $3$ (NS zone). The different symbols (see legenda) are the $\rho_{Sch}$  of GRB$030329$ and GRB$060218$ respectively. The green line is the limit between the black hole ($\rho_{Sch}>1$) and the NS ($\rho_{Sch}\leq1$) zone. It is evident that we got two sets of different possible remnants {\it i.e.} NS or BH.}
\label{Scwarz}
\end{figure}

\paragraph{\bf - How to discriminate between NS or BH solution for the remnant.}
It is worth noticing that the remnant, according to the different models, could be either a millisecond pulsar or a rotating BH as it is foreseen from models (\cite{Meszaros, EkstromS,Yoon2006} and references therein). To ascertain whether the remnant is a BH or a newborn millisecond pulsar like NS (see Fig. \ref{Scwarz}) we used the gravitational parameter $\displaystyle{\rho_{Sch} =\frac{R_{Sch}}{R_{Remn}}}$ {\it i.e.} the ratio between the Schwarzschild radius, $\displaystyle{R_{Sch}=\frac{2 G M}{c^2}}$, and the remnant radius. We used the Schwarzschild radius for the comparison  because the difference with the inner Kerr ergosphere radius is surely within the errors in the determination of remnant mass, angular velocity and radius of the remnant.
%%%%%%%%%%%%%%%%%%%%%%%%%%%%%%%%%%%%%%%%%%%%%%%%%%%%%%%%%
In Fig. \ref{Scwarz} the black  and the blue circles are the results of Model 2 and Model 4 (BH zone) whilst the magenta and red circles are the results of Model $1$ and Model $3$ (NS zone). The different symbols (see legenda)  are the $\rho_{Sch}$  of GRB$030329$ and GRB$060218$ respectively. The green line is the limit between the BH ($\rho_{Sch}>1$) and the NS ($\rho_{Sch}\leq1$) zone. It is evident that we have two sets of different possible remnants NS or BH.
%%%%%%%%%%%%%%%%%%%%%%%%%%%%%%%%%%%%%%%%%%%%%%%%%%%%%%%%
\par
We are also aware that, in our procedure, we treat the emission of GW from a BH in the same way as the emission from a rigid body. Obviously this is not the case, but as far as we are interested in the transition from one state to the other, the assumption can be considered as a limiting case.

\paragraph {\bf - How to discriminate among the possible regions of solutions according to the models.}
In order to probe the detectability of the hypothesized GW signals from progenitors, GRB and remnants, we compare the GW amplitudes obtained for the different models with the sensitivity curves of present and future GW interferometric antennas: VIRGO, LIGO and Advanced LIGO, Advanced VIRGO and Einstein Telescope. We also used those comparisons to try to discriminate among the different models according to the detectability of the GRB and remnants of our sample.

\begin{figure}
\center
\begin{tabular}{|l|}
\hline
\\
\includegraphics[width=12 cm, height=8 cm]{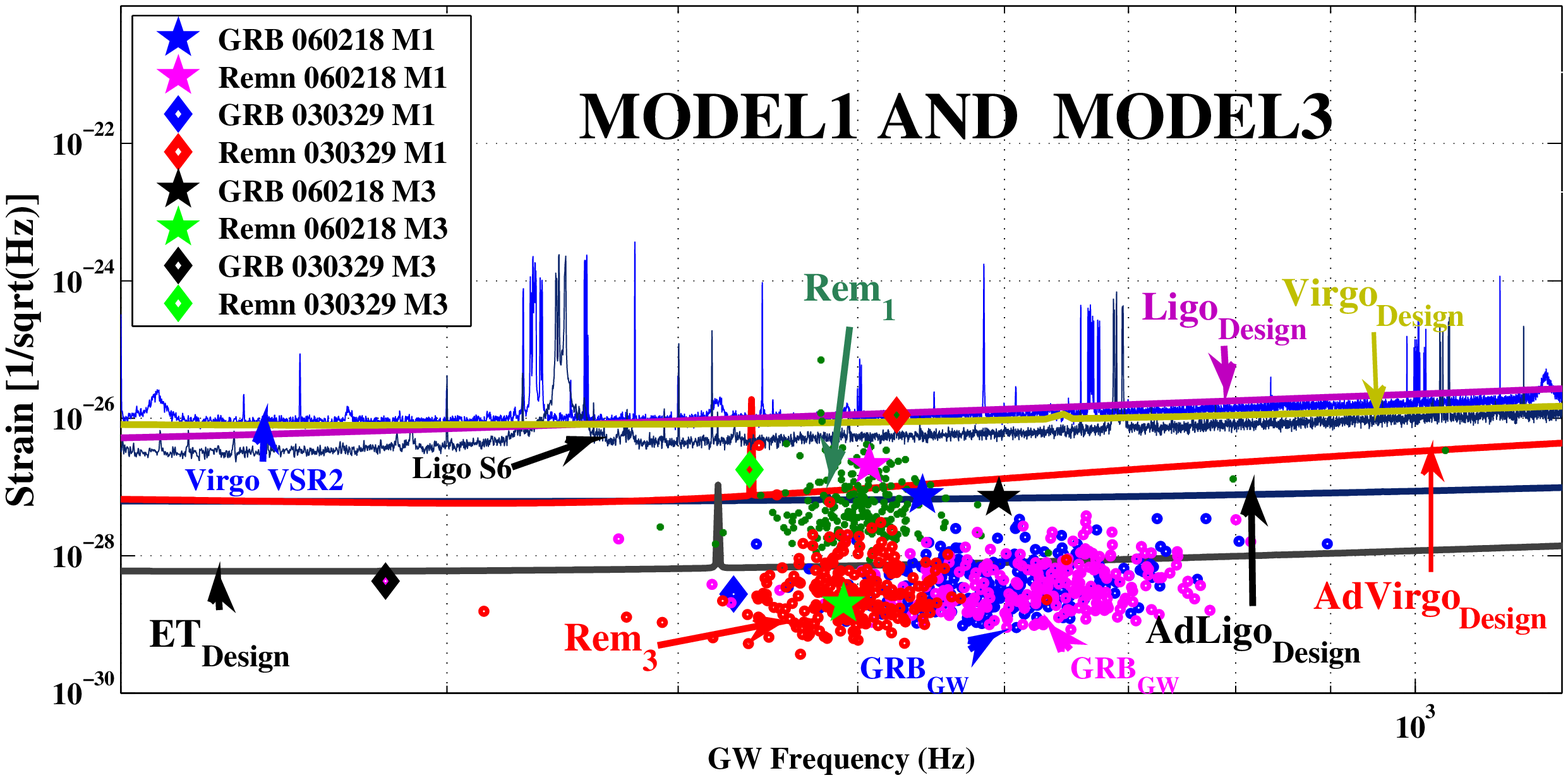}
\\
\includegraphics[width=12 cm, height=8 cm]{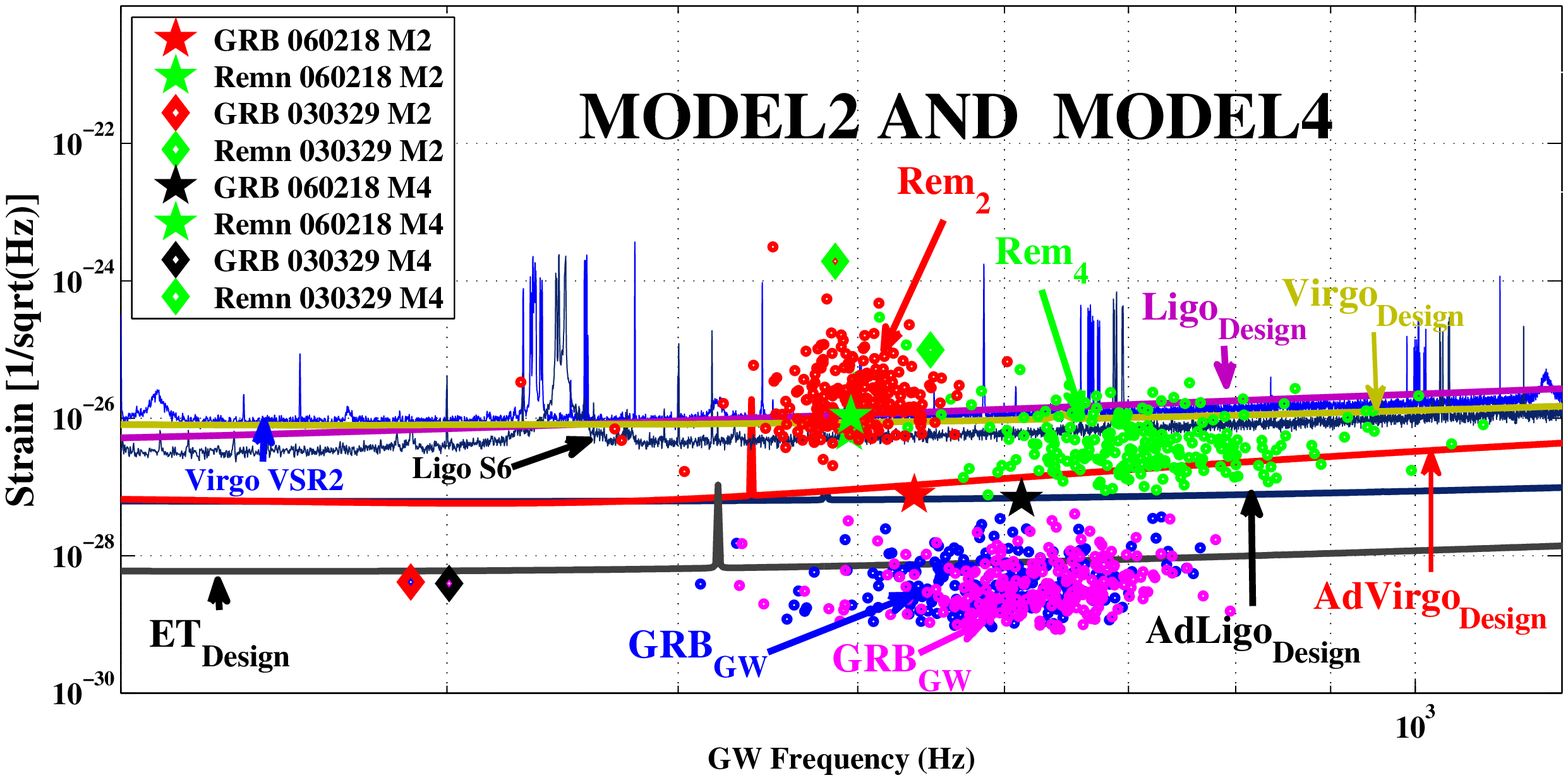}\\
\hline
\end{tabular}
\caption{
On top panel GW amplitudes for the GRB sample are shown for models 1 and 3 (blue and magenta circles respectively) together with GW emission by remnants (NS for these models, red and green circles) and overlapped to  VIRGO and LIGO sensitivity curves assuming the computed $ k_{c} $, {\it i. e.}  energy conversion of $ E^{iso}_{GRB} $ into GW energy.
The GW amplitudes of GRB060218 and GRB030329 are also shown.
On bottom panel the same is shown for models 2 and 4 (BH remnants)
 VIRGO and LIGO performances are shown respectively from the scientific run (VIRGO) $VSR2$ and the scientific $S6$ run (LIGO), compared with the respective target curves \cite{Abbott,Abadie}. An interesting prediction of a remnant millisecond population of either of millisecond pulsar ($Model\,1-Model \,3$: green and red circles top) or  BH ($Model \,2-Model \,4$: red and green circles bottom) can be noticed. }
\label{h_iso}
\end{figure}

In Fig. \ref{h_iso} 
%%%%%%%%%%%%%%%%%%%%%%%%%%%%%%%%%%%%%%%%%%%%%%%%%%%%%%%%%%%%%%%%%%%
on the top  and bottom panels we show the GWs amplitudes we obtained for Model 1 and Model 3 from GRB sample (blue and magenta circles respectively) overlapped to VIRGO and LIGO sensitivity curves assuming the computed $ k_{c} $, {\it i. e.} energy conversion of $ E^{iso}_{GRB} $ into GW energy. The GW amplitudes of GRB$060218$ and GRB$030329$ is also reported on the plots. The inspection of the figure shows that the $ E^{iso}_{GW} $ computed amplitudes, could be observed only with the future facilities specially the Einstein Telescope,  some being at the limit of the foreseen sensitivity for Advanced LIGO, Advanced VIRGO. Specifically, VIRGO and LIGO performances are shown respectively from the scientific run (VIRGO) $VSR2$ and the scientific $S6$ run (LIGO), compared with the respective target curves \cite{Abbott,Abadie}. It is worth  noticing that the amplitudes of $GRB_{GW}$ whatever the Model are practically in the same frequency ranges.  In the figures there is an interesting prediction of a remnant millisecond population of either of millisecond pulsar (Model 1 Model 3: red and green circles top) or  BH (Model 2 and Model 4: red and green circles bottom).
So, according to our approach, we inferred the progenitor mass according to the hypothesis of generation of either a NS or a BH after GRB explosion from a progenitor star, whose type we may infer from evolution star models, using the physical parameters we computed. 
From Fig. \ref{h_iso} it is easy to see that in some cases we can discard the results because, if the model were true we should have already observed GW from the remnants with the present sensitivities of VIRGO and LIGO so we can discard Model 1 Model 2 and Model 4. We can affirm so that, according to the remaining Model 3,  the remnant must be NS and the progenitor masses and radii are in agreement with the hypothesis that the GRB progenitor must be WR stars.
Last but not the least LIGO and VIRGO are presently undergoing to an upgrade to significantly increase their sensitivity, so, the chances to detect a coincident emission of GWs with a GRBs will be substantially improved and the prevision of a millisecond remnant population (NS),   could be experimentally tested also in connection with the estimated conversion energy parameter $k_{c}$.
%%%%%%%%%%%%%%%%%%%%%%%%%%%%%%%%%%%%%%%%%%%%%%%%%%%%%%%%%%%%%%%%%%%%%%%

%%%%%%%%%%%%%%%%%%%%%%%%%%%%%%%%%%%%%%%%%%%%%%%%
%%%%%%%%%%%%%%%%%%%%%%%%%%%%%%%%%%%%%%%%%%%%%%%%%%%%%%%%%%%%%%%%%%%%%%%
%%%%%%%%%%%%%%%%%%%%%%%%%%%%%%%%%%%%%%%%%%%%%%%%%%%%%%%%%
\section{Discussion and conclusions}
%%%%%%%%%%%%%%%%%%

We have  considered the hypothesis  that the energy emitted as gravitational radiation from a core-collapsing system  can be obtained by scaling $E^{iso}_{GRB}$ by a scale factor $k$ ranging  from $10^{-6}$ to $10^{-8}$ thus admitting that there has been a conversion of mass into gravitational energy. From this assumption,  it is possible to estimate the progenitor masses.
%\begin {landscape} 
\begin{table*}
\centering
\caption{ Progenitor mass in $M_{\odot}$, progenitor radius $R_{{PROG}_{\odot}}$. Remnant masses, radii and GW remnant emission frequency of  GRB$030329$ and GRB$060218$ according to the different models}.
\begin{tabular}{|l|l|l|l|l|l|l|}
\hline
{\bf GRB 030329}	&	$M_{PROG}$	&	$ R_{PROG}$	&	 $M_{Remn}$	&	$R_{Remn}$	&	 $f_{Remn} $ \\	
$\pm\sigma$	&	$\pm10 \, M_{\odot}$	&	$\pm5 \, R_{\odot}$	&	$\pm0.8 \, M_{\odot}$	&	$\pm4\, (km)$	&	$\pm280\, Hz$	 \\ \hline
 Modello 1	&	$39$	&	$23$	&	$2$	&	$10$	&	507	 \\ \hline
Modello 2	&	$38$	&	$19$	&	$5$	&	$11$	&	495	 \\ \hline
Modello 3	&	$24   $	&	$10$	&	$2$	&	$10$	&	491	 \\ \hline
Modello 4	&	$24$	&	$9$	&	$7$	&	$12$	&	496	 \\ \hline\hline
{\bf GRB 060218}	&	$M_{PROG}$	&	$R_{PROG}$	&	$M_{RMN} $	&	$R_{RMN}$ 	&	$f_{RMN} $ 	\\
$\pm\sigma$	&	$\pm 8\, M_{\odot}$	&	$\pm 5\, R_{\odot}$	&	$\pm 0.5\, M_{\odot}$	&	$\pm 3 \, km$	&	$\pm 250 \,Hz $ 	 \\ \hline
Model 1	&	27	&	49	&	2	&	10	&	514	 \\ \hline
Model 2	&	24	&	18	&	5	&	15	&	459	 \\ \hline
Model 3	&	21	&	8	&	2	&	8	&	393	 \\ \hline
Model 4	&	21	&	8	&	5	&	15	&	460	 \\ \hline
\end{tabular}
\label{table4}
\end{table*}

\begin{table*}
\centering
\caption{ $k$ of energy conversion, progenitor oblateness $\epsilon_{PRG}$,  progenitor GW frequency	$\omega_{PRG} \, (rad/s)$, remnant	oblateness $\epsilon_{Remn}$ , of GRB$030329$ and GRB$060218$ according to the different models}
\begin{tabular}{|l|l|l|l|l|l|}
\hline
{\bf GRB030329}	&	$k$	&	$ f_{GRB}^{GW}$	&	$\omega_{PROG}$	&	$\epsilon_{PRG}$	&	$\epsilon_{Remn}$	\\ 
$\pm\sigma$	&	$\pm 2.6\times10^{-6}$	&	$\pm270 \,Hz$	&	$\pm4\times10^{-5}\,rad/s$	&	$\pm0.07$	&	$\pm3\times10^{-7}$	\\ \hline
Model 1	&	$5\times10^{-6}$	&	542	&	$5\times10^{-5}$	&	$0.14$	&	$5\times10^{-7}$	\\ \hline
Model 2	&	$5\times10^{-6}$	&	536	&	$5\times10^{-5}$	&	$0.13$	&	$5\times10^{-7}$	\\ \hline
Model 3	&	$5\times10^{-6}$	&	596	&	$8\times10^{-5}$	&	$0.15$	&	$5\times10^{-7}$	\\ \hline
Model 4	&	$5\times10^{-6}$	&	613	&	$8\times10^{-5}$	&	$0.15$	&	$5\times10^{-7}$	\\ \hline\hline
{\bf GRB060218}	&	$k$	&	$ f_{GRB}^{GW}$	&	$\omega_{PROG}$	&	$\epsilon_{PRG}$	&	$\epsilon_{Remn}$	\\ 
$\pm\sigma$	&	$\pm 2.5\times10^{-6}$	&	$\pm260 \,Hz$	&	$\pm1\times10^{-5}\,rad/s$	&	$\pm0.08$	&	$\pm3\times10^{-7}$	\\ \hline
 Modello 1	&	$5\times10^{-6}$	&	387	&	$1.4\times10^{-5}$	&	$0.10$	&	$7\times10^{-7}$	\\ \hline
Modello 2	&	$4\times10^{-6}$	&	400	&	$2.4\times10^{-5} $	&	$0.13$	&	$2\times10^{-6}$	\\ \hline
Modello 3	&	$56\times10^{-6}$	&	383	&	$3.1\times10^{-5} $	&	$0.12$	&	$2\times10^{-5}$	\\ \hline
Modello 4	&	$6\times10^{-6}$	&	427	&	$2.7\times10^{-5}$	&	$0.12$	&	$8\times10^{-7}$	\\ \hline
\end{tabular}
\label{table4_C}
\end{table*}
%%%%%%%%%%%%%%%%%%%%%%%%%%%%%%%%%%%%%%%%%%%%%%%%%%%%%%%%%%%%%%%%%%%%%%%%
From our GRB sample, we analyze the case of GRB$060218$ whose progenitor could be a WR star \cite{Campana} and the case of GRB$030329$ \cite{Goran} and references therein.
We consider these GRB as test cases, because they could constitute an independent support for the newborn magnetars scenario, proposed to account for shallow decays or plateaus \cite{Dai1999,Zhang2002,Fan2006,Yu2007}. The above mentioned support comes from the observation of SN2006aj, associated
with the nearby sub-energetic GRB$060218$, suggesting that the supernova-GRB connection may extend to a much broader range of stellar masses than previously thought, possibly involving two different mechanisms: a 'collapsar' for the more massive stars \cite{Mazzali2002,Mazzali2006a,Mazzali2006b} and a magnetars scenario. In Table \ref{table4}, for GRB$060218$ and GRB$030329$, we show, with their errors, the progenitor masses in $M_{\odot}$, progenitor radii $R_{PROG}$ in $R_{\odot}$,  remnant	 masses, radii and GW remnant  emission frequency of both GRB$030329$ and GRB$060218$ respectively, as inferred from our procedure. In Table \ref{table4_C} we show the $k$ of energy conversion, progenitor oblateness $\epsilon_{PRG}$, progenitor GW frequency	$\omega_{PRG} \, (rad/s)$, remnant oblateness $\epsilon_{Remn}$, of both GRB$030329$ and GRB$060218$ according to the different models. For the case of GRB$060218$ it is easy to see that both the progenitor masses and radii computed according to model $3$ and $4$, are in good agreement with the estimations by \cite{Campana} who states that observations provide strong evidence that the GRB progenitor was a WR star, being the star radius definitely smaller than $5 \times 10^{12}cm$, {\it i.e.} smaller than the radius of the progenitors of type II SNe, like blue supergiants ($4 \times 10^{12}\, cm$ for SN1987A or red supergiants ($3 \times 10^{13}\, cm$).
In Fig. \ref{h_iso} GW amplitudes $h_{iso}$ (see legend) of the remnants of GRB$060218$ according to the different models are placed  over the sensitivity curves of GW interferometric antennas in order to discriminate among the different models on the base of the detectability of the GRB and remnants. This allows to discard all models except Model $3$. Finally we want underline that we are proposing a new method of analysis of GRB progenitor mass and, if the redshift is known from $E_{iso}$ it is possible to infer quickly the mass either of the progenitor or the remnant.

\section*{Acknowledgments}

MDL is supported by INFN ({\it iniziative specifiche} TEONGRAV and QGSKY). LM and FG acknowledge the support of INFN Sez. di Napoli (Virgo Experiment). MGD is grateful for the initial support of the l'Oreal Italia Scholarship and the Polish MNiSW through the Grant N N203 380336 and for the final support of the Boncopompagni-Ludovisi Foundation.


\begin{thebibliography}{99}


\bibitem{Abadie}Abadie J. et al., 2012,  {\it ApJ} {\bf 760}, 12.

\bibitem{Abbott}Abbott B.P. et al., 2010, {\it ApJ} {\bf 715}, 1438.

\bibitem{LIGO} Abramovici A. et. al., 1992, {\it Science}, 256, 325.

\bibitem{Barbier}Barbier D., et al., 2006, GCN Circ. 4780.

\bibitem{VIRGO}Bradaschia C. et. al., 1990, {\it N.I.M. in Phys. Res. A} {\bf 289}, 518.

\bibitem{M_Bresco}Bresco M. et. al., 2005, Bresco M., {\it SOFT COMPUTING} {\bf 9}, pp 525-535.

\bibitem{Campana}Campana S. et al., 2006, {\it Nature} {\bf 442}, 1008.

\bibitem{Cobb}Cobb B. E., Bailyn  C. D., van Dokkum  P. G., \& Natarajan, P., 2006, {\it ApJ}, {\bf 645}, L113.

\bibitem{Corsi}Corsi A., \& Meszaros P.,  2009, {\it ApJ}, {\bf 702}, 1171. 

\bibitem{Crowther2007} Crowther, P. A. 2007, {\it ARA\&A}, {\bf 45}, 177

\bibitem{cucchiara} Cucchiara A.  et al. , 2011, {\it ApJ} {\bf736}, 7.

\bibitem{cusumano}Cusumano G., Moretti A., Tagliaferri G., Kennea J., Burrows D., 2006, {\it GCN Circ.}, 4786.

\bibitem{Dai1999} Dai Z. G. ,  Huang Y. F. \&  Lu T.,1999, {\it Ap. J.} {\bf 520 }, 634.

\bibitem{Dainottia}Dainotti M.G., Petrosian V., Singal J., Ostrowski M.,  2013, {\it Astrophys.J.} {\bf 774}, 157. 

\bibitem{Dainottib}Dainotti M.G.,  Cardone V.F., Piedipalumbo E., Capozziello S., 2013, {\it Mon. Not. Roy. Astron. Soc.} {\bf 436}, 82. 

\bibitem{Dainotti2015}Dainotti M.G., Del Vecchio R., Nagataki S., Capozziello S., 2015 {\it ApJ} {\bf 800}, 31.


\bibitem{Davies2}Davies, M. B., King  A., Rosswog  S., \& Wynn  G. 2002, {\it Ap. J.}, {\bf 579}, L63.

\bibitem{Davies} Davies M.B., Levan A.J., Larsson J., 
King A. R., \&  Fruchter A.S., 2007, {\it AIP Conf.Proc.} 906, 78. 

\bibitem{GCN}GCN. 2007, http://gcn.gsfc.nasa.gov/

\bibitem{swift}Gehrels, N., et al. 2004, {\it ApJ},{\bf  611}, 1005.

\bibitem{Georgy2009}C. Georgy, G. Meynet, R. Walder, D. Folini, and A. Maeder, 2009, {\it A\&A}  {\bf 502}, 611.

\bibitem{GeorgyC}Georgy C. et al. 2012 {\it Astr. \& Astroph.}, {\bf 542}, p. 29 

\bibitem{Glatzel1985}Glatzel et al. 1985,

\bibitem{Goran} G\"{o}ran \"{O}. et al.2008, {\it Mon. Not. Roy. Astron. Soc.}  {\bf 387}, 1227.

\bibitem{Guetta} Guetta D. \& Della Valle M., 2007, {\it ApJ} {\bf 657}, L73.


\bibitem{Fan2006}  Fan, Yi-Zhong \& Dong, Xu,2006, {\it Mon. Not. Roy. Astron. Soc.} {\bf 372}, p. L19-L22

\bibitem{Fryer}Fryer C., Woosley S., \&  Hartmann D., 1999, {\it ApJ} 526, 152.

\bibitem{Fryer2}Fryer C. L., Holz  D. E., \& Hughes S. A.  2002, {\it ApJ}, {\bf 565}, 430.

\bibitem{Fryer1}Fryer C. L., Holz, D. E., \& Hughes, S. A., 2002, {\it ApJ}, {\bf 565}, 430.

\bibitem{Heger}Heger A., Fryer C. L., Woosley S. E., Langer N., \& Hartmann D. H., 2003,
{\it ApJ}, {\bf 591}, 288.

\bibitem{Isaacson}Isaacson R.~A., 1968, {\it Phys. Rev.}, {\bf 166}, 1272.

\bibitem{Kalogera} Kalogera, V.\& Baym G.,1996, {\it  ApJ Letters}, {\bf 470}, L61-64

\bibitem{Kobayashi}Kobayashi S. \& Meszaros P., 2002, {\it Astrophys. J.} {\bf 589}, 861.

\bibitem{Meszaros1}Kobayashi, S., \& Meszaros, P. 2003, {\it ApJ}, {\bf 585}, L89.

\bibitem{Kotake2012} Kotake K., Takiwaki T. and Harikae S. 2012,  {\it Astrophys. J.} {\bf 755}, 84.

\bibitem{Liang}Liang E.W. et al 2006, {\it ApJ} {\bf 653}, L81.

\bibitem{Maggiore} Maggiore M., 2007, {\it Gravitational Waves: Theory and Experiments},  Oxford Univ. Press, Oxford.

\bibitem{Massey2003} Massey, P. 2003, {\it ARA\&A}, {\bf 41}, 15. 

\bibitem{Mazzali2002}Mazzali P.A., 2002,  {\it Astrophys. J.} {\bf 572}, L61.

\bibitem{Mazzali2006a}Mazzali P.A., 2006a,  {\it Astrophys. J.} {\bf 645}, 1323.

\bibitem{Mazzali2006b}Mazzali P.A., 2006b,  {\it Nature} {\bf 442}, 1018.

\bibitem{Milano1997} Milano L., Barone F.,Milano M.,1997, {\it
Phys. Rev. D}, {\bf 55},6 .

\bibitem{Milano2002} Milano M., Koumoutsakos P., 2002 {\it  Journ of Comput. Physics}, {\bf 175}, 79-107.
	
\bibitem{gravitation} Misner C.W., Thorne  K. S., and Wheeler J. A., {\it Gravitation} (Freeman, San Francisco, 1973).

\bibitem{Meszaros} Meszaros P., 2007, {\it Rep. Prog. Phys}. {\bf 69}, 2259.

 \bibitem{MeynetG} Meynet G. et al., 2010, {\it Rev. Mex. de Astron. y Astr. (Serie de Conferencias) } {\bf 38},  113-116 

\bibitem{Ott}Ott, C. D. 2009, {\it Classical and Quantum Gravity}, {\bf 26}, 063001.

\bibitem{VO}Oppenheimer J.R. \&  Volkoff  G.M. , 1939, {\it Phys. Rev. } {\bf 55}, 374.

\bibitem{Osborn}Osborne J. P., Beardmore A. P., Evans P. A., \& Goad M. R., 2011, {\it GCN Circ.}, 11712.

\bibitem{Paczynski}Paczynski B., 1998, {\it Astrophys. J.} {\bf 494}, L45.

	
\bibitem{Petitpas}Petitpas G., Zauderer A., Berger E., Patel N., Brassfield  E., Miller J., Dosaj A., 2011, {\it GRB Coordinates Network, Circular Service}, {\bf 11650}, 1.
 
\bibitem{Petrovic}
 Petrovic J. ,Pols O.,   Langer N.,2006,{\it A\&A}, {\bf 450},219.

\bibitem{Piro}Piro A. L. \& Pfahl E.,  2007, {\it ApJ}, {\bf 658}, 1173.

\bibitem{Price} Price W.L. 1976,{\it Computer J.}, {\bf 7},303

\bibitem {ET}Punturo M. et al.  2010,{\it Class. Quantum Grav.} {\bf 27} 084007.

\bibitem{Romero}Romero G. E., Reynoso  M. M. \& Christiansen H. R. 2010, {\it A\&A}, {\bf 524}, A4.


\bibitem{Sakamoto}Sakamoto, T. et al.,  2005, {\it ApJ}, {\bf 629}, 311. 

\bibitem{Shibata}Shibata  M., Shigeyuki  K., \& Yoshiharu  E., 2003, {\it MNRAS}, {\bf 343}, 619.

\bibitem{Smartt 2009}Smartt, S.J.  2009 {\it ARA\&A}, {\bf 47}, 63.

\bibitem{EkstromS} Ekstr\"{o}m S. et. al. ,2012, {\it Astron. \& Astrophys.},  {\bf 537}, p. 146

\bibitem{vanPutten}van Putten, M. H., et al., 2004, {\it Phys. Rev. D}, {\bf 69}, 044007.

\bibitem{Tominaga2005} Tominaga et al. {\it ApJ}, {\bf 633}, L97.

\bibitem{tanv}Tanvir, N. R. et al.,  2009, {\it Nature} {\bf 461} 1254.

\bibitem{Woosley} Woosley S. E, 1993, {\it ApJ} {\bf 405}, 273.

\bibitem{Woosley2011} Woosley S. E, 2011 arXiv:1105.4193

\bibitem{Woosley1} Woosley, S. E.\& Bloom, J.S. ,  2006, {\it Annual Review of Astronomy and Astrophysics} {\bf 44}, 507.

\bibitem{Yoon2006} Yoon S.C., Langer N., Norman C. 2006,  {\it A \& A} {\bf 460}, 199.

\bibitem{Yu2007} Yu, Y.\& Huang, Y.-F.,2007,{\it Chin. Journ. of Astr. and Astroph.} {\bf 7}, 669

\bibitem{Zhang2002} Zhang B. \& Meszaros P.,2002,{\it Ap. J.} {\bf 566},712-722
\end{thebibliography}
\end{document}